\documentclass[preprint,english,showpacs,preprintnumbers,amsmath,amssymb,floatfix]{revtex4}
\usepackage[T1]{fontenc}
\usepackage[latin9]{inputenc}
\usepackage{color}
\usepackage{array}
\usepackage{amstext}
\usepackage{graphicx}
\usepackage{esint}

\makeatletter

\@ifundefined{textcolor}{}
{%
 \definecolor{BLACK}{gray}{0}
 \definecolor{WHITE}{gray}{1}
 \definecolor{RED}{rgb}{1,0,0}
 \definecolor{GREEN}{rgb}{0,1,0}
 \definecolor{BLUE}{rgb}{0,0,1}
 \definecolor{CYAN}{cmyk}{1,0,0,0}
 \definecolor{MAGENTA}{cmyk}{0,1,0,0}
 \definecolor{YELLOW}{cmyk}{0,0,1,0}
 }

\@ifundefined{definecolor}
 {\usepackage{color}}{}
\@ifundefined{definecolor}
 {\usepackage{color}}{}
\makeatother

\makeatother

\usepackage{babel}

\begin{document}

\title{The Role of Different Schemes in the QCD Analysis and Determination of the Strong Coupling}

\author{A.~Vafaee}
\email[]{A.vafaee@semnan.ac.ir}
\author{A.~N.~Khorramian}
\email[]{Khorramiana@semnan.ac.ir}
%\thanks{}
\affiliation{Faculty of Physics, Semnan University, P. O. Box 35131-19111, Semnan, Iran}

\date{\today}

\begin{abstract}
In this article, we present a Next-to-Leading Order (NLO) QCD analysis to study the role and influence of different schemes on simultaneous determination of the Parton Distribution Functions (PDFs) and strong coupling, $\alpha_s(M^2_Z)$. We perform our analysis based on three different data sets, HERA I and II combined data, H1-ZEUS charm combined data, and  H1 and ZEUS  beauty production cross sections data, in two different Thorne-Roberts (TR or RT) and Thorne-Roberts Optimal (RT OPT) schemes. We show in going from RT scheme to RT OPT scheme, in addition of reduction the uncertainty of some PDFs, specially for the gluon distribution, we get $\sim 0.4$~\% and $\sim 0.7$~\%~improvement in the fit quality and $\sim 0.9$~\% and $\sim 1.6$~\%~improvement for the strong coupling, $\alpha_s(M^2_Z)$, without and with heavy flavor contributions, respectively.  
\end{abstract}

%\pacs{13.60.Hb, 12.39.-x, 14.65.Bt ???}
%\keywords{}

\maketitle

\section{\label{introduction}Introduction}
In perturbative Quantum Chromodynamics (pQCD), the proton structure is described in terms of the parton density functions, $f(x)$, which is the probability of finding a parton, either gluon or quark, with a fraction $x$ of the proton's momentum. This probability depends on the factorization scale, $\mu_{\rm f}^2$, the scale at which the proton structure is probed, which for inclusive Deep Inelastic Scattering (DIS) is usually taken as $Q^2$.
It is customary to present these functions as parton momentum distributions, $xf(x)$, and are called parton distribution functions (PDFs).  To calculate cross section for $e^{\pm}p$, $pp$ and $p{\overline{p}}$ colliders, these parton distribution functions are convoluted with the fundamental point-like 
scattering cross sections for partons.

The proton PDFs are extracted classically from QCD fits by a measure of the agreement between experimental data and theoretical models. The PDFs are then evolved using coupled integro-differential Dokshitzer-Gribov-Lipatov-Altarelli-Parisi (DGLAP)~\cite{DGLAP} evolution equations, at  the leading order (LO), next-to-leading order (NLO) and next-to-next-to-leading order (NNLO). However, pQCD does not predict the parton distribution functions at the initial scale and  they need to be determined by fits to the experimental data.

 One of the important subject in QCD analysis and PDF fits, is attention to the theoretical description of heavy quarks and number of
 active flavors such as charm and bottom. The approach of heavy flavors in proton structure functions has an important effect on determination of PDFs obtained in fits to proton structure function and consequently on the predictions for cross sections at various hadron colliders such as the Tevatron and LHC.  
 
 The combination of 
 HERA I and II data  at HERA has been recently reported in  Ref.~\cite{Abramowicz:2015mha}. These data with heavy quark production cross sections provide  an important constraint on PDFs, particularly on gluon PDF at low $x$. Several different theoretical groups are provided the PDF sets using HERA fixed target and hadron-collider experimental data  such as MSTW \cite{Martin:2009iq}, HERAPDF \cite{Aaron:2009aa}, CTEQ/CT \cite{Pumplin:2002vw,Lai:2010vv}, 
ABM \cite{Alekhin:2008hc,Alekhin:2009vn,Alekhin:2012ig}, NNPDF \cite{Ball:2008by,Mironov:2009uv},  and  JR \cite{Jimenez-Delgado:2014twa}.  
 
 There are various schemes for separating in the proton structure functions into calculable processes and PDFs. Investigate the role and influence of two different  RT (Thorne-Roberts) and RT OPT (Thorne-Roberts Optimal) schemes on simultaneous determination of PDFs, fit quality and strong coupling, $\alpha_s(M^2_Z)$, is the main topic in this analysis. We perform our analysis based on three different data sets, HERA I and II combined data~\cite{Abramowicz:2015mha}, charm quark cross section H1-ZEUS combined data~\cite{Abramowicz:1900rp} and  H1 and ZEUS  beauty production cross sections data~\cite{Aaron:2009af,Abramowicz:2014zub}, in two different Thorne-Roberts~\cite{Thorne:2006qt} and Thorne-Roberts Optimal~\cite{Thorne:2012az} schemes and compare the central values of PDFs and their uncertainties, fit quality and numerical values of strong coupling with each other. 
 
 The paper is structured as follows. In Sec.~II, we describe the theoretical base of DIS as a powerful tool for probing the proton structure and also introduce the reduced  $e^{\pm}p$ scattering cross sections.  We introduce the functional form and PDF parametrization  in Sec.~ III. In Sec.~IV, we introduce some different schemes and discuss specially, about two different RT and RT OPT schemes. In Sec.~V, we determine our QCD analysis PDFs and discuss about the role and influence of two different RT and RT OPT schemes on the fit quality. In Sec.~VI, we investigate the role and influence of two different RT and RT OPT schemes on determination of PDFs and strong coupling, $\alpha_s(M^2_Z)$. Finally in Sec.~VII, we present our discussion and conclusion.

\section{Cross sections in DIS}
Deep inelastic electron (positron) scattering on proton at centre-of-mass energies of up to $\sqrt{s} \simeq 320\,$GeV
at HERA play a central role to the exploration
of proton structure and quark--gluon dynamics as
described by perturbative quantum chromodynamics. A large phase space in Bjorken scale, $x$~, and negative of four-momentum-transfer squared, $Q^2$, have been explored at HERA. The reduced Neutral Current (NC) deep inelastic $e^{\pm}p$ scattering cross sections can be expressed in terms of generalized structure functions by
\begin{eqnarray}
   \sigma_{r,NC}^{{\pm}}&=&   \frac{d^2\sigma_{NC}^{e^{\pm} p}}{d{x}dQ^2} \frac{Q^4 x}{2\pi \alpha^2 Y_+} =\tilde{F_2} \mp \frac{Y_-}{Y_+} x\tilde{F_3} -\frac{y^2}{Y_+} \tilde{F_{\rm L}}~,
    \label{eq:NC}
\end{eqnarray} 
where  $Y_{\pm} = 1 \pm (1-y)^2$ and $\alpha$ is the fine-structure constant which is defined 
at zero momentum transfer. Here, the generalized structure functions, $\tilde F_2$, $\tilde F_L$ and $\tilde F_3$ may be expressed as a linear combinations of five structure functions of proton, $F^{\gamma}_2, F^{\gamma Z}_2, F^{\gamma Z}_3, F^Z_2$ and $F^Z_3$, relating to pure photon exchange, photon--$Z$ interference and pure $Z$ exchange, 
respectively.  These structure functions are depend
on the electroweak parameters as  \cite{Beringer:1900zz}
\begin{eqnarray} \label{strf}                                                   
 \tilde F_2 &=& F_2 - \kappa_Z v_e  \cdot F_2^{\gamma Z} +                      
  \kappa_Z^2 (v_e^2 + a_e^2 ) \cdot F_2^Z~, \nonumber \\   
 \tilde F_L &=& F_{\rm L} - \kappa_Z v_e  \cdot F_{\rm L}^{\gamma Z} +                      
  \kappa_Z^2 (v_e^2 + a_e^2 ) \cdot F_{\rm L}^Z~, \nonumber \\                     
 x\tilde F_3 &=&  - \kappa_Z a_e  \cdot xF_3^{\gamma Z} +                    
  \kappa_Z^2 \cdot 2 v_e a_e  \cdot xF_3^Z~,                                   
\end{eqnarray}
where $v_e$ and $a_e$ are the vector and axial-vector weak couplings of the electron to the $Z$ boson, and $\kappa_Z(Q^2)$, defined as $\kappa_Z(Q^2) = Q^2 /[(Q^2+M_Z^2)(4\sin^2 \theta_W \cos^2 \theta_W)]$, with $\theta_W$ as Weinberg angle. In this article, we perform QCD fit analysis using  xFitter \cite{xFitter} open source framework, which perviously was known as HERAfitter. In xFitter, the values of $Z$-boson mass and the electroweak mixing angle are, $M_Z=91.1876$\,GeV and $\sin^2 \theta_W=0.23127$, respectively.

 At low value of $Q^2$, $Q^2\ll M_Z^2$, the $Z$ exchange contribution may be ignored and therefore the reduced NC DIS cross sections can be expressed by 
\begin{eqnarray}
%  \nonumber
   \sigma_{r,NC}^{{\pm}}&= &   F_2  - \frac{y^2}{Y_{+}}  F_L~.
    \label{eq:NClast}
\end{eqnarray}

 Similarly, the inclusive unpolarized  CC $e^{\pm}p$  scattering reduced cross sections given by
\begin{eqnarray}
%\centering
\sigma_{r,CC}^{\pm} &=&\frac{2\pi x}{G^2_F} \left[\frac{M^2_W+Q^2}{M^2_W}\right]^2 \frac{d^2\sigma_{CC}^{e^{\pm} p}}{d{x}dQ^2} =\frac{Y_{+}}{2}  W_2^{\pm} \mp \frac{Y_{-}}{2}x  W_3^{\pm} - \frac{y^2}{2} W_L^{\pm}~,
\end{eqnarray}
where, $\tilde W_2^{\pm}$, $\tilde W_3^{\pm}$ and $\tilde W_L^{\pm}$  are  another set of structure functions and $G_F$ is the Fermi constant, given by $G^2_F = {e^2}/[{4\sqrt{2}~{\sin ^2\theta_W}M^2_W}]$, with $e$, as electromagnetic coupling constant.
%= {g^2}/[{4M_W}]$
In xFitter QCD framework, the values of  $M_W=80.385$~GeV and $G_F=1.16638\times 10^{-5} $~GeV$^{-2}$
were used for the $W$-boson mass and Fermi constant.

 In the Quark Parton Model (QPM), $W_{\rm L}^\pm = 0$ and depending on the 
charge of the lepton beam,  $W_2^\pm$, $xW_3^\pm$ structure functions may be represent by sums and differences of quark and anti-quark distributions, $i. e.$   $W_2^{+}  \approx  x\overline{U}+xD$, 
    $W_2^{-}  \approx  xU+x\overline{D}$, 
   $xW_3^{+}  \approx   xD-x\overline{U}$, and 
   $xW_3^{-}  \approx  xU-x\overline{D}$. The  $xU$, $x \overline{U}$ and $xD$, $x \overline{D}$ terms, denote the sums of $u$ and $d$-type quarks and anti-quarks distributions, respectively. The mentioned sums are related to the quark distributions as $xU=xu+xc$,  $x \overline{U}=x\overline{u}+x\overline{c}$ and $xD=xd+xs$,  $x \overline{D}=x\overline{d}+x\overline{s}$  in below the bottom quark  mass threshold. It is clear that by assuming symmetry between
the quarks and anti-quarks for sea PDFs, the valence PDFs can be expressed as  $xu_v=xU-x\overline{U}$  and $xd_v=xD-x\overline{D}$

Accordingly, the CC $e^+p$ and $e^-p$ cross sections, at the Leading Order (LO) are sensitive to  the different combinations of the quark flavor densities, as follow:
\begin{eqnarray}
%\centering
\sigma_{r,CC}^{+} \approx  x\overline{U}+ (1-y)^2xD =
       x [\overline u + \overline c] + (1-y)^2 x [d+s]~, \\
\sigma_{r,CC}^{-} \approx  xU +(1-y)^2 x\overline{D}=x[u+c] + (1-y)^2 x[\overline d + \overline s]~.
\end{eqnarray}

The reduced cross sections for heavy-quark production, $\sigma_{red}^{Q\bar{Q}}$ ($Q=b,c$), in analogy to the inclusive NC deep inelastic $e^{\pm}p$ scattering cross section, may be expressed by
\begin{eqnarray}
	\sigma_{red}^{Q\bar{Q}} &=& 
	          \frac{d\sigma^{Q\bar{Q}}(e^{\pm} p)}{d{x} \, dQ^2} \cdot \frac{x \, Q^4}{2 \pi \alpha^2 Y_{+}} =F_2^{Q\bar Q} \mp \frac{Y_{-}}{Y_{+}}x  F_3^{Q\bar Q} - \frac{y^2}{Y_{+}}  F_L^{Q\bar Q}~, 
    \label{eq:NCheavy}
\end{eqnarray}	
where $\alpha$ is the electromagnetic coupling constant, $Y_{\pm} = (1 \pm (1-y)^2)$ and $F_2^{Q\bar{Q}}$, $xF_3^{Q\bar{Q}}$ and $F_L^{Q\bar{Q}}$ are heavy-quark contributions to the 
inclusive structure functions $F_2$, $xF_3$ and $F_L$, respectively.

 In the kinematic region at HERA, the $F_2^{Q\bar{Q}}$ structure function makes a dominant contribution. The $xF_3^{Q\bar{Q}}$ structure function makes the contribution only from $Z^0$ exchange and $\gamma Z^0$ and which implies for $Q^{2} \ll M_{Z}^{2}$ region, this contribution can be ignored. Finally, the contribution of longitudinal heavy-quark structure function, $F_L^{Q\bar{Q}}$, is suppressed only for  $y^2 \ll 1$ region which may be a few percent in the kinematic region accessible at HERA and therefore can not be ignored. Therefore, neglecting the $xF_3^{Q\bar Q}$ structure function contribution, the reduced heavy-quark cross section, $\sigma_{red}^{ Q\bar Q}$, for both positron and electron beams may be expressed by  
\begin{eqnarray}
%  \nonumber
   \sigma_{red}^{ Q\bar Q}&=&   \frac{d^2\sigma^{Q\bar Q}(e^{\pm}p)}{dxdQ^2} \frac{xQ^4}{2\pi\alpha^2 Y_{+}} =F_2^{Q\bar Q}  - \frac{y^2}{Y_{+}}  F_L^{Q\bar Q}~. 
    \label{eq:NCheavylast}
\end{eqnarray}
Accordingly, at high $y$, the reduced charm-quark cross section, $\sigma_{red}^{ Q\bar Q}$, and $F_2^{Q\bar Q}$ structure function only differ by a small $F_L^{Q\bar Q}$ contribution \cite{Daum:1996ec}.
\section{Data sets and PDF parametrization}
In this analysis, we use full seven sets of HERA I and II combined NC and CC DIS $e^{\pm}p$ scattering cross sections \cite{Abramowicz:2015mha} data, as our central data set, along with charm quark cross section H1-ZEUS combined data~\cite{Abramowicz:1900rp} and  H1 and ZEUS  beauty production cross sections data~\cite{Aaron:2009af,Abramowicz:2014zub}. Based on these recently reported data sets, we perform four different fits at next-to-leading order to study the role and influence of different schemes on simultaneous determination of the PDFs QCD-fit quality and strong coupling, $\alpha_s(M^2_Z)$. The availability of these new experimental combined inclusive deep inelastic $e^{\pm}p$ scattering data over a large phase space in $x$ and $Q^2$, allows us to make new proton PDFs independent of any nuclear or deuterium corrections.  

 For NC and CC $e^{\pm}p$ scattering, the combined reduced cross sections, depend on the centre-of-mass energy, $\sqrt{s}$, and further, on the two kinematic variables $x$ and $Q^2$ . 
The kinematic variable $x$, in turn, is related to $y$,  $Q^2$ and $s$ through the relationship $x=Q^2/(sy)$. The kinematic ranges for NC cross sections data are: $0.045 \leq Q^2 \leq 50000 $\,GeV$^2$
and  $6 \cdot 10^{-7} \leq x \leq 0.65$ 
at values of the inelasticity, $y = Q^2/(sx)$, 
between $0.005$ and $0.95$. Also, the kinematic ranges for CC cross sections data are: $200 \leq Q^2 \leq 50000 $\,GeV$^2$ and $1.3 \cdot 10^{-2} \leq x \leq 0.40$ at values of the inelasticity $y$ between $0.037$ and $0.76$.  

 For NC and CC cross sections, the total uncertainties are below 1.5\,\% over the $Q^2$ range of $3 \le  Q^2 \le 500$\,GeV$^2$ and below 3\,\% up to  $Q^2 = 3000$\,GeV$^2$. The proton beam energies: $E_p$ = 460, 575, 820 and 920 GeV which are corresponding to $\sqrt{s} \simeq$ 225, 251, 300 and 320 GeV and the invariant mass of the hadronic system, $W$, for these events have a minimum  
of $15$\,GeV.

 However, at high $Q^2$, some differences between the reduced NC $e^+p$ and $e^-p$ scattering cross sections, together with the high-$Q^2$ CC data, constrain the valence-quark distributions. Without taking account strong isospin symmetry, as done in the deuterium data analysis, the CC $e^+p$ data, particularly constrain the valence down-quark distribution in the proton. The lower-$Q^2$ NC data, constrain the low-$x$ sea-quark distributions. These data, also, constrain the gluon distribution through their precisely measured $Q^2$ variations.  The inclusion of NC data at different beam energies, such that the $\tilde{F_{\rm L}}$ is probed through the $y$ dependence of the cross sections, make a further constraint on the gluon distribution.
 
 The charm quark with pole mass of $m_c=1.5$ GeV is a heavy quark which is accessible kinematically at HERA and measurements of charm production cross sections in deep inelastic $e^{\pm}p$ scattering at HERA from the H1 and ZEUS Collaborations are combined. For charm production reduced cross section measurements data \cite{Abramowicz:1900rp} the kinematic range of Bjorken scaling variable and and photon virtuality are $3\cdot 10^{-5} \le x\le 5\cdot 10^{-2}$ and $2.5\le Q^2 \le 2000$ GeV$^2$, respectively.
 
  The beauty quark with pole mass of $m_b=4.75$ GeV is a heavy quark which is accessible kinematically at HERA from H1 and ZEUS. For H1, the reduced cross sections for beauty production are obtained in the kinematic range of Bjorken scaling variable $2\cdot 10^{-4} \le x \le 5\cdot 10^{-2}$ and photon virtuality $5.0\le Q^2 \le 2000$ GeV$^2$ and for ZEUS, the reduced cross sections for beauty production are obtained in the kinematic range of Bjorken scaling variable $2\cdot 10^{-4} \le x \le 5\cdot 10^{-2}$ and the photon virtuality $5.0\le Q^2 \le 1000$ GeV$^2$.
  
  To include the heavy-flavor contributions, we use two different schemes, the Thorne-Roberts~\cite{Thorne:2006qt} and Thorne-Roberts Optimal~\cite{Thorne:2012az} schemes and then compare the central values of PDFs, fit quality and numerical values of strong coupling, $\alpha_s(M^2_Z)$ with each other. In our methodology we choose $\mu_f = \mu_r=Q$~, as a perturbative quantum chromodynamics scales with pole masses $m_b=4.75$ GeV and $m_c=1.5$ GeV.

 In this analysis based on the HERAPDF approach \cite{Abramowicz:2015mha}, we generically parameterized the PDFs of the proton, $xf(x)$, at the initial scale of the QCD evolution $Q^2_0= 1.9$ GeV$^2$ as
\begin{equation}
 xf(x) = A x^{B} (1-x)^{C} (1 + D x + E x^2)~,
\label{eqn:pdf}
\end{equation}
where in the infinite momentum frame, $x$ is the fraction of the proton's momentum. To determine the normalization constants $A$ for the valence and gluon distributions, we use the QCD number and momentum sum rules.  Using the functional form of PDFs in above equation, one can consider the central parametrisation of PDFs for $xu_v$ and $xd_v$ valence quark distributions, and 
the  $x\bar{U}(x)$ and $x\bar{D}(x)$ anti-quark distributions for $u$ and $d$-type at the starting scale of $Q_0^2$:
\begin{eqnarray}
\label{eq:xuvpar}
xu_v(x) &=  & A_{u_v} x^{B_{u_v}}  (1-x)^{C_{u_v}}\left(1+E_{u_v}x^2 \right) , \\
xd_v(x) &=  & A_{d_v} x^{B_{d_v}}  (1-x)^{C_{d_v}} , \\
x\bar{U}(x) &=  & A_{\bar{U}} x^{B_{\bar{U}}} (1-x)^{C_{\bar{U}}}\left(1+D_{\bar{U}}x\right) , \\
x\bar{D}(x) &= & A_{\bar{D}} x^{B_{\bar{D}}} (1-x)^{C_{\bar{D}}}~.
\end{eqnarray}

As we mentioned  before, $x\bar{U}(x)=x\bar{u}(x)$ and  $x\bar{D}=x\bar{d}+x\bar{s}$ at the initial scale of $Q_0^2$. Also in this scale, the strange quark distribution may expressed of down sea quarks such as $x \bar s= f_s x \bar D$  \cite{Abramowicz:2015mha,Martin:2009iq,Nadolsky:2008zw} which $f_s$ is fixed to 0.31$\pm$0.08 in xFitter. By setting
$A_{\bar{U}}=A_{\bar{D}} (1-f_s)$ and the requirement $B_{\bar{U}}=B_{\bar{D}}$ can be imposed a further constraint to ensure that
$x\bar{u} \rightarrow x\bar{d}$ as $x \rightarrow 0$ can be imposed additional constraints. Therefore, sea distribution for down and strange quraks will be $x\bar{d}=(1-f_s)x\bar D$ and $x\bar{s}=f_s x\bar D$ respectively.

The gluon PDF functional form given by
\begin{eqnarray}
xg(x) &=   & A_g x^{B_g} (1-x)^{C_g} - A_g' x^{B_g'} (1-x)^{C_g'} ~.
\label{eq:xgpar}
\end{eqnarray}
The gluon functional form, $xg(x)$, is an exception from Eq.~(\ref{eqn:pdf}), because of extra subtracted term of the form $A_g' x^{B_g'} (1-x)^{C_g'}$.
Really, the behavior of the PDFs for low $x$ and high $x$ values can control by $x^B$ and $(1-x)^{C}$, respectively.  According to Ref.~\cite{Abramowicz:2015mha,Martin:2009iq}, the above extra term can control the low $x$, where  the single $x^{B_g}$ term can not control the gluon  behavior at very low $x$. As suggested in Ref.~ \cite{Martin:2009iq}, we fixed the value of $C_g'$ and set $C_g'=25$ such that the extra term in the gluon distribution does not contribute at large $x$.

The three normalization parameters $A_{u_v},~A_{d_v}$ and $A_g$ are determined by the QCD sum rules, $i.~e.$ quark number sum rules and also momentum  sum rule. Considering all together this analysis performed by fitting the remaining the 14 free parameters in Eqs.~(\ref{eq:xuvpar}--\ref{eq:xgpar}). By taking into account the strong coupling, $\alpha_s(M^2_Z)$ as an extra free parameter in our QCD analysis, we have 15 unknown parameters which can be extracted using the fits of the data. 

\section{Heavy-quark  schemes}
In this section, we discuss the heavy-flavor corrections to introduce the heavy-quark structure functions which described by Wilson coefficients.
Deep inelastic scattering nucleon structure functions can be explain using different theoretical schemes. In the Fixed Flavor Number (FFN) scheme the charm mass effects are included to a fixed
perturbative order in QCD. For $Q^2\sim m_h^2~(h=c,b)$, the heavy flavors are
created in the final state and described using FFN scheme. In FFN scheme, the heavy-quark masses treated explicitly and the structure function given by 
\begin{equation}
F(x,Q^2)=C^{\rm FF, n_f}_k(Q^2/m_h^2)\otimes f^{n_f}_k(Q^2)~,
\end{equation}
where $n_f$ is the number of light quark flavors. In this scheme, full NLO calculations of heavy-flavor production, exist for DIS
\cite{Ball:2011mu,Gluck:2008gs,Laenen:1992cc,Harris:1997zq,Martin:2006qz,Martin:2010db,Lai:1999wy}. The heavy flavors in this scheme, are considered as massive at all scales, and do not appear as an active flavor within the proton. When all heavy flavors are considered as massive, the number of light flavors in the PDFs is therefore, fixed to 3 and beauty as well as charm are always produced in the matrix element. Generally, at high scales, $Q^2 \gg m_h^2$, heavy flavor behave like 
massless partons and the distributions of different light quark number are related to each other by the perturbative expression:
\begin{equation}
f^{n_f+1}_j(\mu_F^2)= A_{jk}(\mu_F^2/m_H^2)\otimes f^{n_f}_k(\mu_F^2)~,
\label{eq:pdfplus1}
\end{equation}
where the matrix elements, $A_{jk}(\mu_F^2/m_H^2)$, contain the fixed-order $\ln(\mu_F^2/m_h^2)$ contributions.

In the $Q^2/m_h^2 \to \infty$ limit, the FFN scheme becomes the Zero-Mass Variable Flavor Number (ZM-VFN)
scheme. This scheme neglects power suppressed terms in the charm mass. In this case the structure function is written as 
\begin{equation}
F(x,Q^2) = C^{\rm ZMVF,n_f+m}_j\otimes f^{n_f+m}_j(Q^2)~,
\end{equation}
where $m$ is the number of heavy flavor which have effectively become light quarks. In ZM-VFN scheme, the heavy quark mass is set to zero for the computation of the kinematics and
matrix elements. This scheme has been used for most
NLO variable-flavor parton-density fits such as 
NNPDF2.0 \cite{Ball:2010de}, CTEQ6M \cite{Pumplin:2002vw}, H1 \cite{Aktas:2006hy} and ZEUS-S \cite{Chekanov:2002pv}.

The general-mass variable flavor number (GM-VFN) scheme is defined similarly to the ZM-VFN scheme. So, the structure function is given by 
\begin{equation}
F(x,Q^2) = C^{\rm GMVF,n_f+m}_j(Q^2/m_h^2)\otimes f^{n_f+m}_j(Q^2)~.
\end{equation}
Now, the coefficient functions are dependent on $Q^2/m_h^2~(h=c,b)$ and reduce to the ZM-VFNS limit, as $Q^2/m_h^2 \to \infty$. In the GM-VFN scheme, we may rewrite the structure function, $F(x,Q^2)$, by
\begin{eqnarray}
F(x,Q^2) &=& C^{\rm GMVF,n_f+1}_j(Q^2/m_h^2)\otimes f^{n_f+1}_j(Q^2) \nonumber \\
&=& C^{\rm GMVF,n_f+1}_j(Q^2/m_h^2)\otimes A_{jk}(Q^2/m_h^2)\otimes f^{n_f}_k(Q^2) \nonumber \\ 
&\equiv & C^{\rm FF, n_f}_k(Q^2/m_h^2)\otimes f^{n_f}_k(Q^2)~,
\end{eqnarray}
when we consider the transition from $n_f$ active flavors to $n_f+1$.
Accordingly, the GM-VFN scheme may be defined at all orders, from the formal equivalence of the $n_{f}$ flavor and $n_f+1$ flavor descriptions as follow:
\begin{equation}
C^{\rm FF,n_f}_k(Q^2/m_h^2) \equiv 
C^{\rm GMVF,n_f+1}_j(Q^2/m_h^2)\otimes A_{jk}(Q^2/m_h^2)~,
\label{GMVFNSdeffull}
\end{equation}
where for simplicity we set $\mu_F^2=Q^2$. This fact that Eq.(\ref{eq:pdfplus1}) converts $n_f$ flavor PDFs to $n_f+1$ flavor PDFs, ensure us at limit $Q^2/m_h^2 \to \infty$, where all power-suppressed $m_h^2/Q^2$ corrections become ignorable, the GM-VFN scheme coefficient functions become identical to that of ZM-VFN scheme.

 In order to explain how the GM-VFN scheme approach works, we see at ${\cal O}(\alpha_S)$, Eq.~(\ref{GMVFNSdeffull}) 
may be equivalently written as
\begin{eqnarray}
C^{\rm FF,n_f,(1)}_{2,hg}(Q^2/m_h^2) = 
C^{\rm GMVF,n_f+1,(0)}_{2, h\bar h}(Q^2/m_h^2)\otimes P^0_{qg}\ln(Q^2/m_h^2)+
C^{\rm GMVF,n_f+1,(1)}_{2,hg}(Q^2/m_h^2),
\label{GMVFNSdef1}
\end{eqnarray}
which defines the GM-VFN scheme coefficient functions.  
As we mentioned, the coefficient functions must tend to the massless limits as $Q^2/m_h^2 \to \infty$, and Eq.~({\ref{GMVFNSdef1}}) is satisfy this condition.
However, the GM-VFN scheme coefficient functions, $C^{\rm GMVF}_j(Q^2/m_h^2)$ is only uniquely defined in this limit.

 We should note that the freedom to modify the GM-VFN scheme coefficient functions, $C^{\rm GMVF}_j(Q^2/m_h^2)$, by power suppressed terms, while this modification is simultaneously applied to the corresponding subtraction terms, occurs individually in each of structure function or cross-section. We can even change the 
GM-VFN scheme for only, $F_2(x,Q^2)$ while leaving the one for $F_L(x,Q^2)$ the same. As a consequence of our freedom to define various definitions for GM-VFN scheme, leads to the existence of different instructions of GM-VFN scheme which have been reported such as RT~\cite{Thorne:2012az}, 
ACOT~\cite{Aivazis:1993pi,Kramer:2000hn,Tung:2001mv}, FONLL~\cite{Forte:2010ta}, or BMSN~\cite{Buza:1996wv}. A recent discussion of the application of different schemes to heavy flavor data at HERA is reported in Ref.~\cite{Behnke:2015qja}. A comparison between VFN schemes for charm quark  electroproduction is also reported in \cite{Chuvakin:1999nx}.

 The Thorne-Roberts, (TR) scheme \cite{Thorne:2006qt,Thorne:2012az}, which sometimes referred as RT-scheme is a GM-VFN scheme. Some groups such as HERAPDF \cite{Aaron:2009aa}, MSTW \cite{Martin:2009iq}, CT (CTEQ) \cite{Lai:2010vv} and NNPDF \cite{Ball:2008by,Mironov:2009uv}, use GM-VFN schemes in PDF analysis. We should note that, the Thorne-Roberts scheme provides a smooth transition from the massive FFN scheme \cite{Martin:2006qz}, at low scales $Q^2<m_h^2$ to the massless ZM-VFN scheme at high scales $Q^2>>m_h^2$. However, the connection is not unique. A GM-VFN scheme may be defined by demanding equivalence of the $n_{f} = n$ (FFN) and $n_{f} = n +1$ flavor (ZM-VFN) descriptions above the transition point for the new parton distributions (they are by definition identical below this point), at all orders.
 
 Two different variants of the RT schemes are available: RT standard~\cite{Thorne:2006qt} as used in MSTW PDF sets and RT optimal~\cite{Thorne:2012az}, with a smoother transition across the heavy quark mass scales. In addition, using the k-factor technique, two fast version schemes, RT FAST and RT OPT FAST, are available, corresponding to RT (Thorne-Roberts) and RT OPT (Thorne-Roberts Optimal) schemes, respectively. The k-factors are defined as the ratio between massless and massive scheme. They are applied to the fast massless scheme accessed by QCDNUM \cite{Botje:2010ay}. However, the k-factors are only calculated correctly for the PDF parameters which enter the first iteration of the minimization and are not updated with each iteration. Hence the RT FAST and RT OPT FAST calculations must be repeated by inputting the final PDF parameters  and iterating this procedure until the input and output PDFs are not significantly different.

\section{Determination of the PDF Fits}
Evolution of PDFs and determination of unknown parameters based on the two different RT and RT OPT schemes, is the next step in our QCD analysis. 

 Generally, determination of the proton patron distribution functions is a complex attempt involving several steps, specially when we take the strong coupling, $\alpha_s(M^2_Z)$ as a free extra parameter. In this article we use QCDNUM \cite{Botje:2010ay} version 17-01/12 to evolve the PDFs and set the theory type based on DGLAP~\cite{DGLAP} collinear evolution equations. We perform four different fits at the next-to-leading order and set the evolution starting scale, $Q_0^2=$ 1.9 GeV$^2$. As we mentioned, the different heavy flavour schemes are used by different theory groups, we use two different, RT and RT OPT schemes and use HERAPDF as a PDF-Style. The minimization is the next step of our QCD-fit analysis. We use MINUIT~\cite{James:1975dr} program, as a powerful package for minimization, parameter errors and correlations. 

 To determine PDFs unknown parameters, we minimize the  $\chi^2$ function, when we take into account both correlated and uncorrelated measurement uncertainties. The $\chi^2$ function defined by

\begin{equation}
\chi^2=\sum_{i=1}^{N_{\rm pts}}\left(\frac{D_i+\sum_{k=1}^{N_{\rm corr}}
r_k\sigma_{k,i}^{\rm corr}-T_i}{\sigma_i^{\rm uncorr}}\right)^2+\sum_{k=1}^{N_{\rm corr}}r_k^2,
\label{eq:chi2}
\end{equation}
where $D_i+\sum_{k=1}^{N_{\rm corr}}r_k\sigma_{k,i}^{\rm corr}$ are the data values allowed to shift by some multiple $r_k$ of the systematic error, $\sigma_{k,i}^{\rm corr}$, to give the best 
fit result, and $T_i$ are the parametrized predictions.
%\clearpage

\begin{table}[h]
\begin{center}
\begin{tabular}{|l|c|c|c|c|}
\hline
\hline
{ \bf Order} & \multicolumn{4}{c|}{ {\bf NLO} }    \\ \hline
 { \bf Experiment} & {$~~~~$RT BASE$~~~~$} & {RT OPT BASE} & { $~~~$RT TOTAL$~~~$} & { RT OPT TOTAL} \\ \hline
  HERA I+II CC $e^{+}p$ \cite{Abramowicz:2015mha} & 45 / 39& 45 / 39& 45 / 39& 45 / 39 \\ 
  HERA I+II CC $e^{-}p$ \cite{Abramowicz:2015mha} & 49 / 42& 49 / 42& 49 / 42& 49 / 42 \\ 
  HERA I+II NC $e^{-}p$ \cite{Abramowicz:2015mha} & 222 / 159& 222 / 159& 221 / 159& 222 / 159 \\ 
  HERA I+II NC $e^{+}p$ 460 \cite{Abramowicz:2015mha} & 209 / 204& 210 / 204& 209 / 204& 210 / 204 \\ 
  HERA I+II NC $e^{+}p$ 575 \cite{Abramowicz:2015mha} & 213 / 254& 212 / 254& 214 / 254& 212 / 254  \\
  HERA I+II NC $e^{+}p$ 820 \cite{Abramowicz:2015mha} & 66 / 70& 66 / 70& 66 / 70& 66 / 70 \\ 
  HERA I+II NC $e^{+}p$ 920 \cite{Abramowicz:2015mha} & 422 / 377& 418 / 377& 424 / 377& 419 / 377 \\ \hline
 {Charm H1-ZEUS} \cite{Abramowicz:1900rp} & - & - & 40 / 47& 39 / 47  \\   
{H1 beauty} \cite{Aaron:2009af} & - & - & 2.0 / 12& 3.4 / 12   \\ 
{{ZEUS beauty}} \cite{Abramowicz:2014zub} & - & - & 11 / 17& 13 / 17 \\ \hline
 { Correlated ${\bf \chi^2}$} & 109& 108& 125& 118 \\\hline
{\bf {Total $\bf{\chi^2}$ / dof}}  & ${\bf \frac{1335}{1130}}$ &  ${\bf \frac{1330}{1130}}$  & ${\bf \frac{1406}{1206}}$ &  ${\bf \frac{1396}{1206}}$  \\ \hline
\hline
    \end{tabular}
\vspace{-0.0cm}
\caption{\label{tab:data}{Data sets used in our NLO QCD analysis, with corresponding partial $\chi^2$ per data point for each data set including  $\chi^2$ per degrees of freedom (dof) for two different RT and RT OPT schemes.}}
\vspace{-0.4cm}
\end{center}
\end{table}

As we mentioned, we perform our analysis based on three different data sets, HERA I and II combined data~\cite{Abramowicz:2015mha}, charm quark cross section H1-ZEUS combined data~\cite{Abramowicz:1900rp} and  H1 and ZEUS  beauty production cross sections data~\cite{Aaron:2009af,Abramowicz:2014zub}. Now to be clear, we sometimes refer to HERA run I and II combined data as ``BASE'' and BASE plus all other remaining data sets as ``TOTAL''. The total number of data points for BASE and TOTAL data sets are 1307 and 1388, respectively. On the other hand, we perform this QCD analysis with $Q^2 \geq {Q^2_{\rm min}=3.5}$~GeV$^2$ cut and this cut on $Q^2$, reduces the total number of data points from 1307 to 1145 for BASE and from 1388 to 1221 for TOTAL data sets, as can be seen from Table~\ref{tab:data}. Now based on Table \ref{tab:data}, we may present our QCD fit quality for HERA I and II combined data only and for  RT and RT OPT schemes:

\begin{eqnarray}
\frac{\chi^2_{\rm TOTAL}}{dof} &=& \frac{1335}{1130}=1.181~~{\rm for~ RT~BASE~,} \\
\frac{\chi^2_{\rm TOTAL}}{dof} &=& \frac{1330}{1130}= 1.176~~\rm for~ RT~OPT~BASE~, 
\end{eqnarray}

Also our QCD fit quality for HERA I and II combined data with heavy quark production cross sections data  for  RT and RT OPT schemes as:
\begin{eqnarray}
\frac{\chi^2_{\rm TOTAL}}{dof} &=& \frac{1406}{1206}=1.165~~\rm for~ RT~TOTAL~, \\
\frac{\chi^2_{\rm TOTAL}}{dof} &=& \frac{1396}{1206}=1.157~~\rm for~ RT~OPT~TOTAL~.
\end{eqnarray}
As we can  deduce from Eqs.~(23-26), we obtain four different values of $\chi^2_{\rm TOTAL}$/dof, corresponding to four different fits, which in turn imply four different fit-quality in some PDFs. Since the relative change in a quantity such as $\chi^2$ is defined by $\frac{\bigtriangleup \chi^2}{\chi^2_{\rm RT}}$, with  $\bigtriangleup \chi^2=\chi^2_{\rm RT}-\chi^2_{\rm RT\;OPT}$, so according to Table \ref{tab:data}, in going from RT scheme to RT OPT scheme, we get $\sim 0.4$~\% and $\sim 0.7$~\%~improvement in the fit quality, without and with the heavy flavor contributions included, respectively. As we explain in the next section, this differences of fit quality, implies a significance reduction of some PDFs uncertainties, specially for gluon distributions and some of it's ratios.   

\section{Investigate the Role and Influence of Different Schemes}
Now, we present the role and influence of different schemes on simultaneous determination of parton distribution functions and strong coupling, $\alpha_s(M^2_Z)$. Also, we present our numerical fit results for the PDFs, $\alpha_s(M^2_Z)$  and their uncertainties at the next-to-leading order for two different RT and RT OPT schemes.

 We perform this analysis based on two separate scenarios. In the first scenario, we fix $\alpha_s(M^2_Z)$ to 0.117, as a default value for strong QCD scale parameter in the xFitter framework, and make our QCD fit analysis based on only 14 unknown free parameters, according to Eqs.~(\ref{eq:xuvpar}--\ref{eq:xgpar}). Although in this scenario, we obtain the four different values of $\chi^2_{\rm TOTAL}$ / dof, but we find nothing to show the role and influence of different schemes  on determination of parton distribution functions and their uncertainties. In the second scenario we consider the strong coupling, $\alpha_s(M^2_Z)$ as a free parameter and try to determine it's value by refit our data sets, but this time with 15 unknown free parameters. Based on second scenario, not only we obtain, as previous, the four different values of $\chi^2_{\rm TOTAL}$ / dof value, but also, as we expected, we clearly find the influence of different schemes on PDFs, specially on gluon distribution and some of it's ratios. As we know the strong coupling constant, $\alpha_s(M^2_Z)$, play a central role in pQCD factorization theorem and the result of this analysis emphasis on it's dramatic correlation with PDFs. From this point of view, we can say the strong coupling, $\alpha_s(M^2_Z)$, play a central role to reveal the impact of different schemes on determination of PDFs,  in this analysis.

 In Table \ref{tab:parh}, we present a next-to-leading order numerical values of parameters and their uncertainties for  the 
$xu_v$, $xd_v$, sea and gluon PDFs at the input scale of $Q^2_0 = 1.9$~GeV$^2$ for two different RT and RT OPT schemes. 
 
 As we mentioned, the strong coupling, $\alpha_s(M^2_Z)$, when considered as a free parameter, play a central role to reveal the influence of different schemes on determination of PDFs. According to Table \ref{tab:parh}, we obtain {${ \alpha_s(M^2_Z)}$} $=$ ${0.1161 \pm 0.0043}$ and ${0.1151 \pm 0.0032}$ for RT BASE and RT OPT BASE, respectively and also we obtain {${ \alpha_s(M^2_Z)}$} $=$ ${ 0.1177 \pm 0.0039}$ and ${0.1158 \pm 0.0028}$ for RT TOTAL and RT OPT TOTAL, respectively. These values may be compared with world average values $\alpha_s(M^2_Z)=0.1185 \pm 0.0006$ reported by the Particle Data Group (PDG) \cite{Agashe:2014kda}. According  to the relative change in the strong coupling,  $\frac{\bigtriangleup \alpha_s(M_Z^2)}{\alpha_s(M_Z^2)_{\rm RT}}$, with  $\bigtriangleup \alpha_s(M_Z^2)=\alpha_s(M_Z^2)_{\rm RT}-\alpha_s(M_Z^2)_{\rm RT\;OPT}$, now according to Table \ref{tab:parh}, in going from RT scheme to RT OPT scheme we get $\sim 0.9$~\% and $\sim 1.6$~\%~improvement in the central values of the strong coupling, $\alpha_s(M^2_Z)$, without and with heavy flavor contributions included, respectively. 
 
 In Table \ref{tab:final}, we compare the numerical values of $\frac{\chi^2_{\rm TOTAL}}{dof}$ and ${\alpha_s(M^2_Z)}$ for two different RT and RT OPT schemes.    

\begin{table}[h]
\begin{center}
\begin{tabular}{|l|c|c|c|c|}
\hline
\hline
 \multicolumn{5}{|c|}{ {\bf NLO} }    \\ \hline
 { \bf Parameter} & {$~~~~$RT BASE$~~~~$} & {RT OPT BASE} & { $~~~$RT TOTAL$~~~$} & { RT OPT TOTAL} \\ \hline
  ${B_{u_v}}$ & $0.712 \pm 0.046$& $0.710 \pm 0.045$& $0.723 \pm 0.046$& $0.712 \pm 0.043$ \\ 
  ${C_{u_v}}$ & $4.88 \pm 0.11$& $4.89 \pm 0.11$& $4.83 \pm 0.11$& $4.86 \pm 0.10$ \\ 
  $E_{u_v}$ & $13.9 \pm 2.6$& $13.7 \pm 2.2$& $13.5 \pm 2.6$& $13.6 \pm 2.2$  \\ \hline
  ${B_{d_v}}$ & $0.811 \pm 0.093$& $0.812 \pm 0.093$& $0.824 \pm 0.094$& $0.816 \pm 0.093$  \\ 
  $C_{d_v}$ & $4.18 \pm 0.42$& $4.24 \pm 0.38$& $4.17 \pm 0.42$& $4.21 \pm 0.38$ \\ \hline
  $C_{\bar{U}}$ & $9.1 \pm 1.1$& $9.21 \pm 0.87$& $8.67 \pm 0.96$& $8.89 \pm 0.82$  \\ 
  $D_{\bar{U}}$ & $18.5 \pm 4.2$& $19.2 \pm 3.9$& $16.2 \pm 3.6$& $17.5 \pm 3.5$ \\ 
  $A_{\bar{D}}$ & $0.160 \pm 0.013$& $0.158 \pm 0.010$& $0.161 \pm 0.013$& $0.1607 \pm 0.0100$ \\ 
  $B_{\bar{D}}$ & $-0.166 \pm 0.012$& $-0.1728 \pm 0.0083$& $-0.166 \pm 0.012$& $-0.1709 \pm 0.0080$ \\ 
  $C_{\bar{D}}$ & $4.4 \pm 1.3$& $4.4 \pm 1.3$& $4.5 \pm 1.3$& $4.6 \pm 1.4$ \\\hline
  $B_g$ & $-0.13 \pm 0.19$& $-0.10 \pm 0.12$& $-0.12 \pm 0.21$& $-0.11 \pm 0.12$  \\ 
  $C_g$ & $11.8 \pm 3.5$& $13.5 \pm 2.3$& $10.5 \pm 2.7$& $12.5 \pm 2.0$ \\ 
  $A_g'$ & $2.3 \pm 2.1$& $3.4 \pm 1.7$& $1.8 \pm 1.7$& $2.9 \pm 1.2$ \\ 
  ${B_g'}$ &  $-0.217 \pm 0.093$& $-0.164 \pm 0.096$& $-0.217 \pm 0.097$& $-0.179 \pm 0.096$ \\  
  \hline 
  {${ \alpha_s(M^2_Z)}$} & ${0.1161 \pm 0.0043}$& ${0.1151 \pm 0.0032}$ & ${0.1177 \pm 0.0039}$& ${0.1158 \pm 0.0028}$ \\ \hline
\hline
    \end{tabular}
\vspace{-0.0cm}
\caption{\label{tab:parh}{ {The NLO numerical values of parameters and their uncertainties for  the 
$xu_v$, $xd_v$, $x\bar u$, $x\bar d$, $x\bar s$ and $xg$ PDFs at the initial scale of $Q^2_0 = 1.9$~GeV$^2$, for two different RT and RT OPT schemes.}}}
\vspace{-0.4cm}
\end{center}
\end{table}

\begin{table}[t]
\scriptsize
\begin{ruledtabular}
\begin{tabular}{lcc}
% {Order}& \multicolumn{3}{c}{\ \ \ \ {\bf NLO} } \\ \hline
 {Scheme} & $\chi^2_{\rm TOTAL}/dof$ & ${\alpha_s(M^2_Z)}$ \\ \hline 
  {RT BASE}  & {${1.181}$}& ${ 0.1161 \pm 0.0043}$  \\ 
  {RT OPT BASE}  & {${ 1.176}$}& ${ 0.1151 \pm 0.0032}$  \\ 
  {RT TOTAL} & {${ 1.165}$}& ${ 0.1177 \pm 0.0039}$  \\ 
  {RT OPT TOTAL} & {${ 1.157}$}& ${ 0.1158 \pm 0.0028}$  \\
\end{tabular}
\caption{\label{tab:final}{Comparison of the numerical values of $\frac{\chi^2_{\rm TOTAL}}{dof}$ and ${\alpha_s(M^2_Z)}$  for two different RT and RT OPT schemes \cite{Thorne:2006qt,Thorne:2012az} .}}
\end{ruledtabular}
\end{table}

\section{Discussion and Conclusion}
In this paper, we present a next-to-leading order, QCD analysis to study the role and influence of different schemes on simultaneous determination of the PDFs and strong coupling, $\alpha_s(M^2_Z)$.  Also in the current study, we show the central role of the strong coupling, $\alpha_s(M^2_Z)$ in reveal of the impact of heavy flavor contribution in different schemes, when we considered it as a free parameter which should be determined through the fit process.

We perform our analysis based on three different data sets, HERA I and II combined data, H1-ZEUS combined data and  H1 and ZEUS  beauty production cross sections data, in two different RT and RT OPT schemes. In the first scenario, which $\alpha_s(M_Z^2)$ is fixed to $0.117$,  all the input parameters are almost unchanged by varying from RT-scheme and RT OPT scheme. In fact, based on 14 unknown free parameters, in the RT/RT OPT scheme alone, when we added the heavy flavors data we find no reduction of the gluon or other PDFs uncertainties. At the second scenario, when we considered strong coupling, $\alpha_s(M_Z^2)$,  as an extra free parameter, we find clearly heavy flavor impacts on the parton distribution functions, specially on the gluon distribution and some of it's ratios, both in the RT and RT OPT schemes. Obviously, if we would like to investigate the role and influence of two different RT and RT OPT schemes in the QCD analysis with and without heavy flavors contribution, firstly we need to reveal the impact of heavy flavor contributions on the PDFs distributions and from this point of view we may say the strong coupling, $\alpha_s(M_Z^2)$, play a central role in this regard. We show in going from RT scheme to RT OPT scheme, we get $\sim 0.9$~\% and $\sim 1.6$~\%~improvement in the central value of the strong coupling, $\alpha_s(M^2_Z)$, and we get $\sim 0.4$~\% and $\sim 0.7$~\%~improvement in the fit quality, without and with heavy flavor contributions included, respectively.

\begin{figure*}
\includegraphics[width=0.32\textwidth]{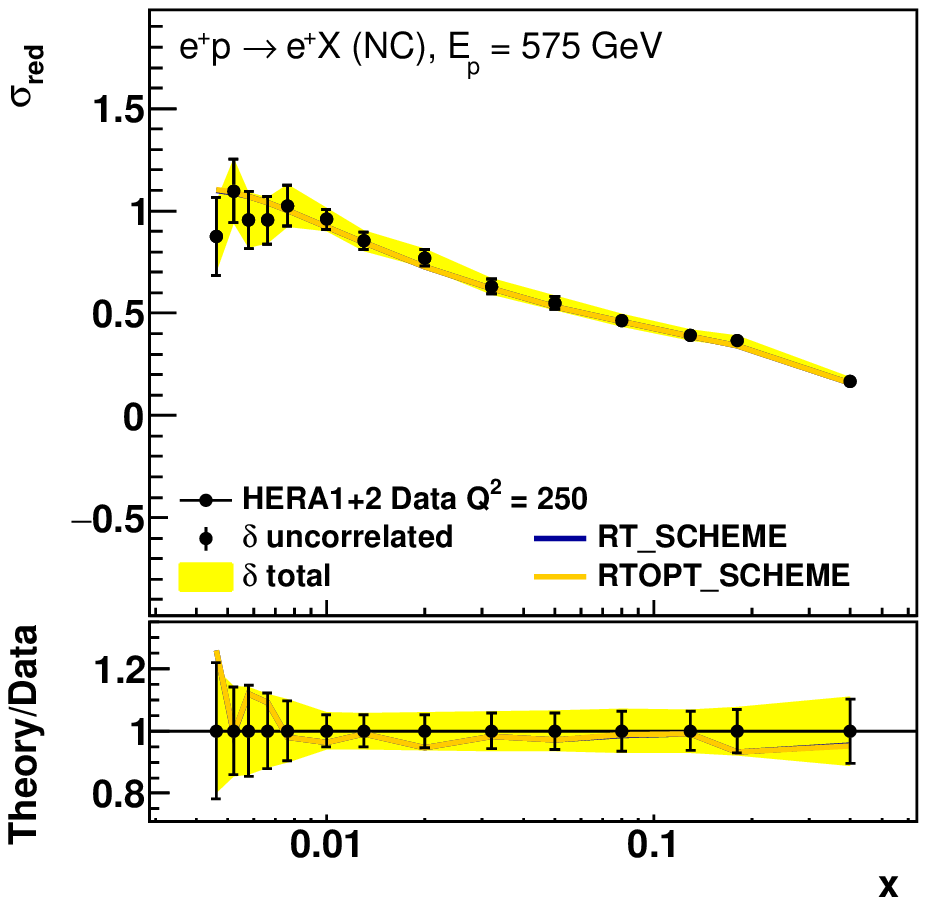}
\includegraphics[width=0.32\textwidth]{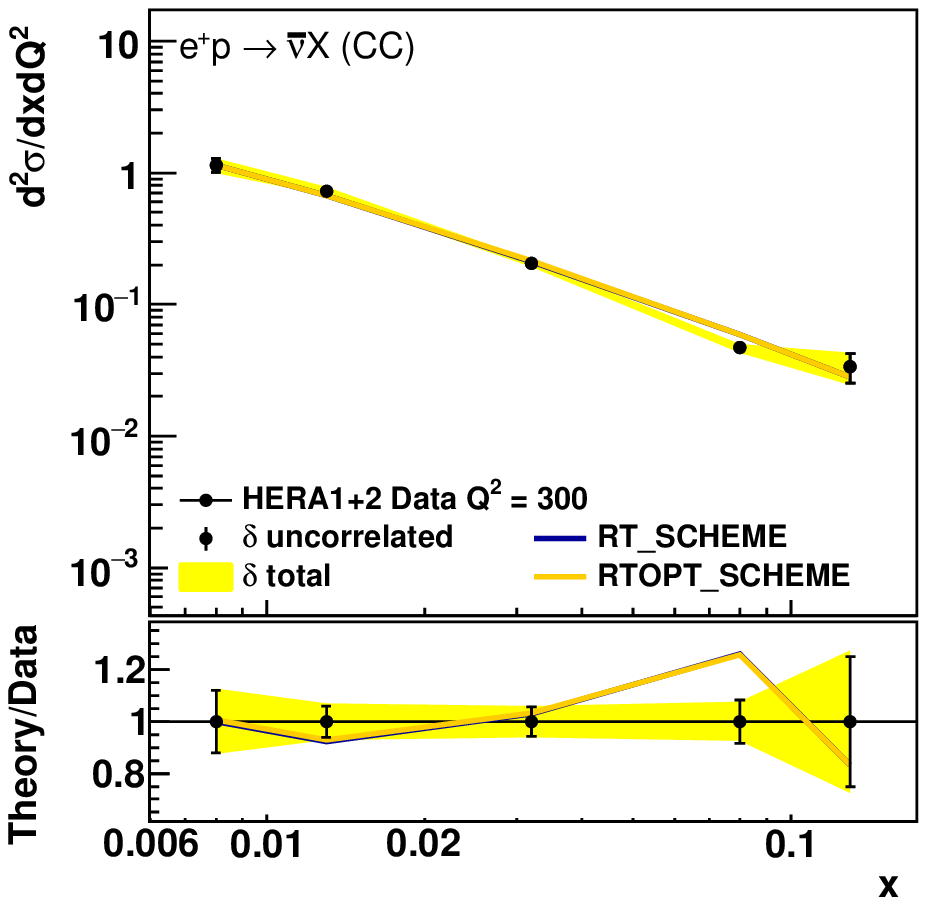}
\includegraphics[width=0.32\textwidth]{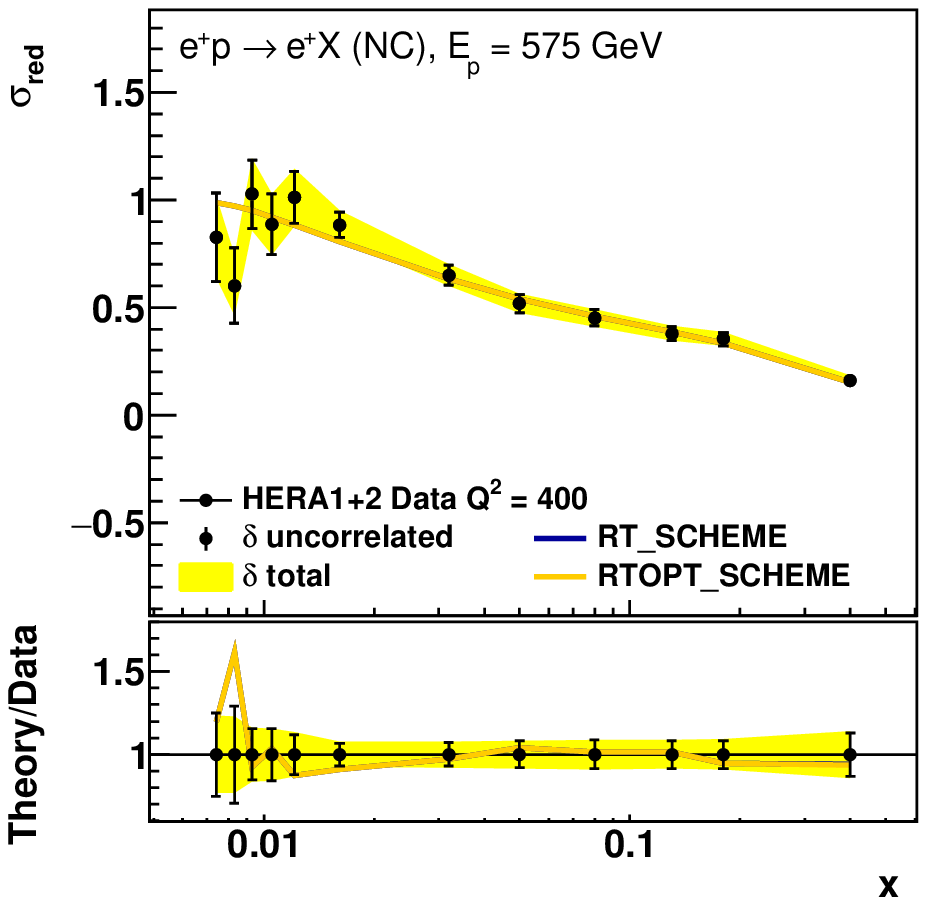}

\includegraphics[width=0.32\textwidth]{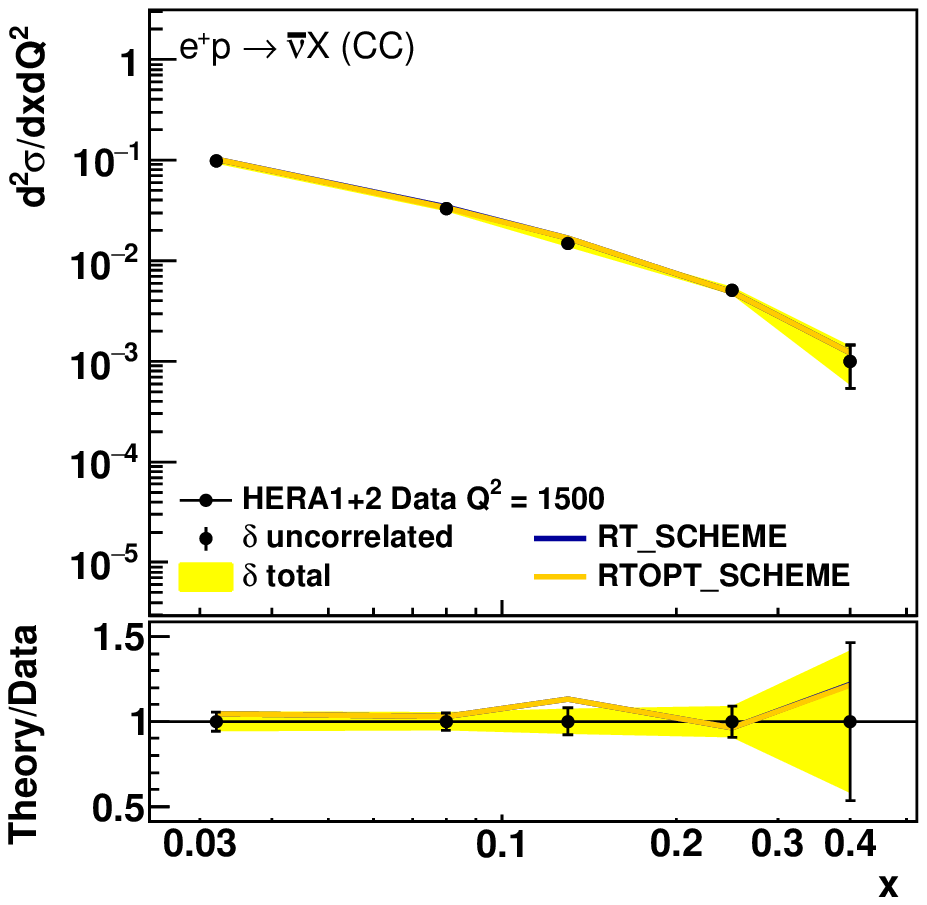}
\includegraphics[width=0.32\textwidth]{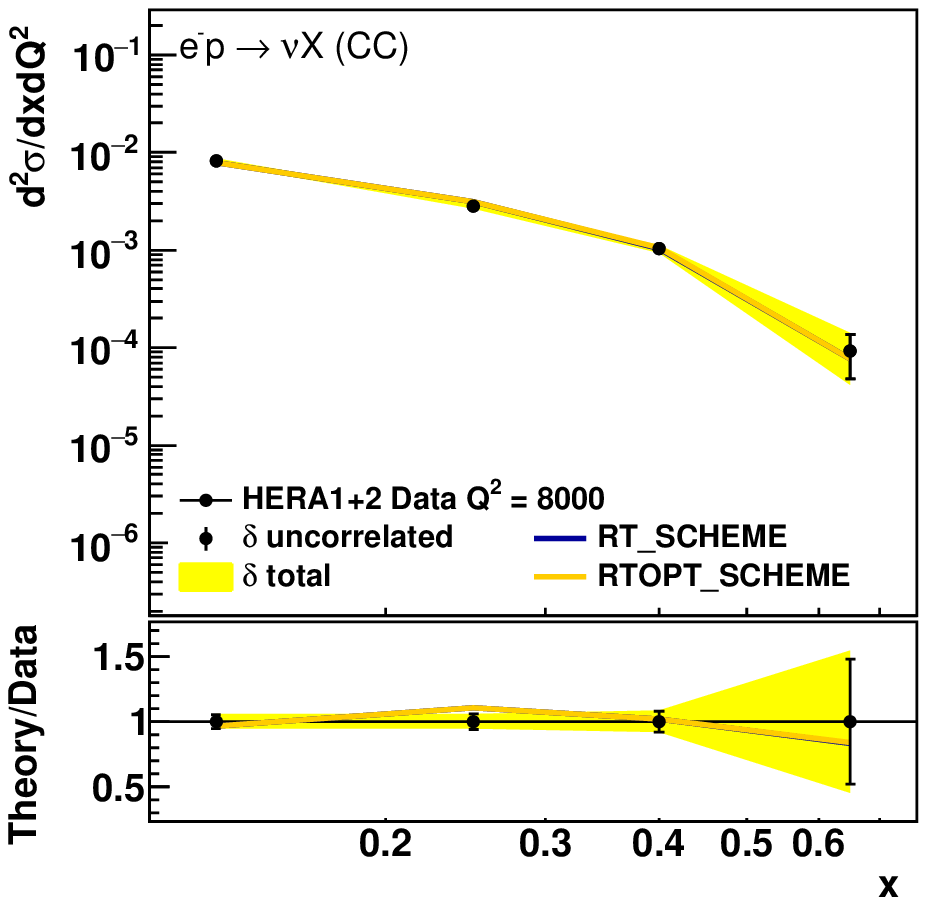}
\includegraphics[width=0.32\textwidth]{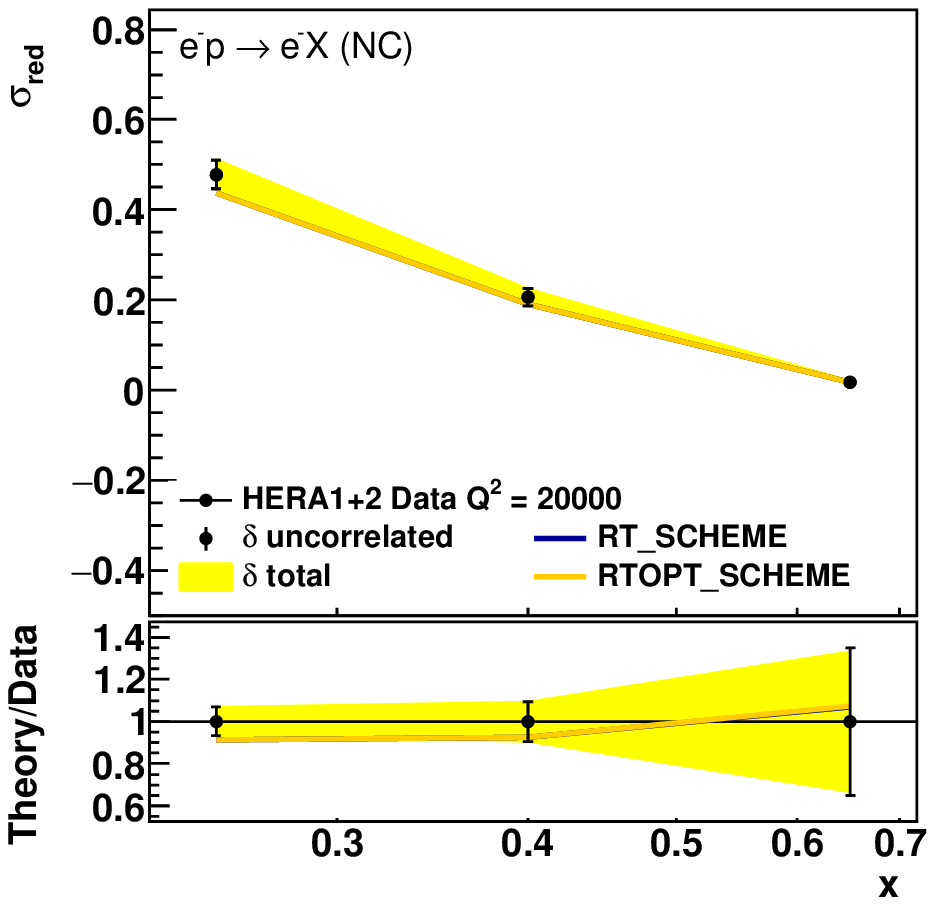}

\includegraphics[width=0.32\textwidth]{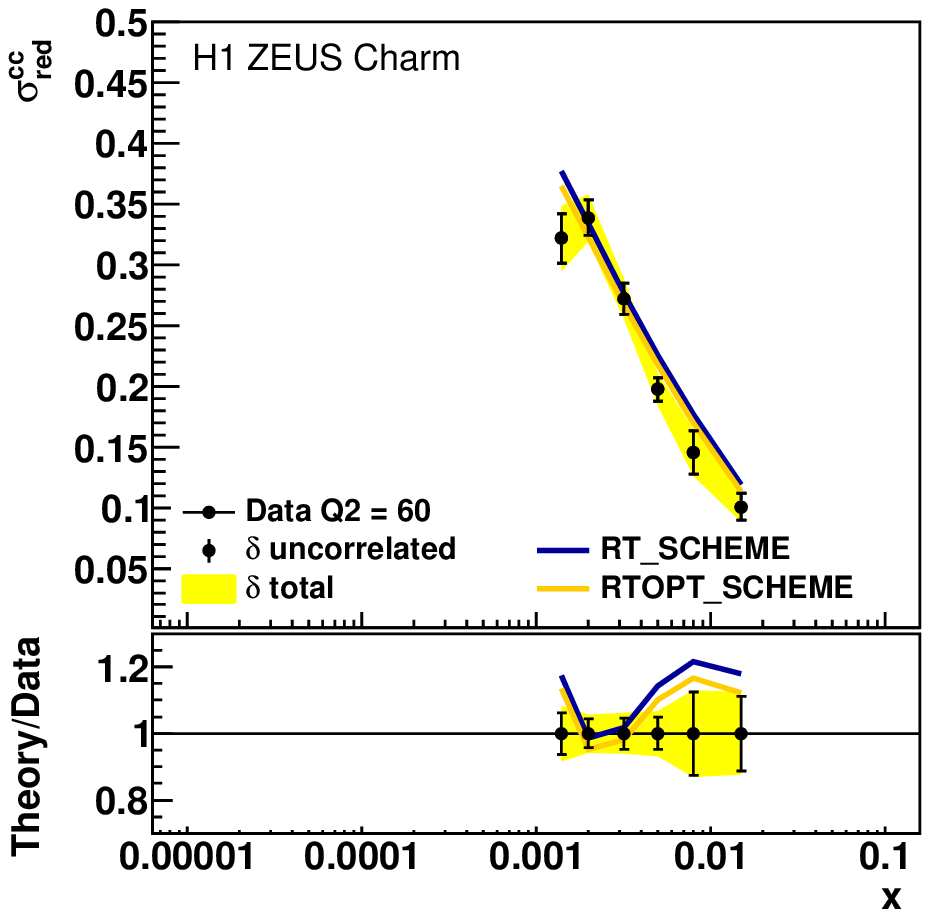}
\includegraphics[width=0.32\textwidth]{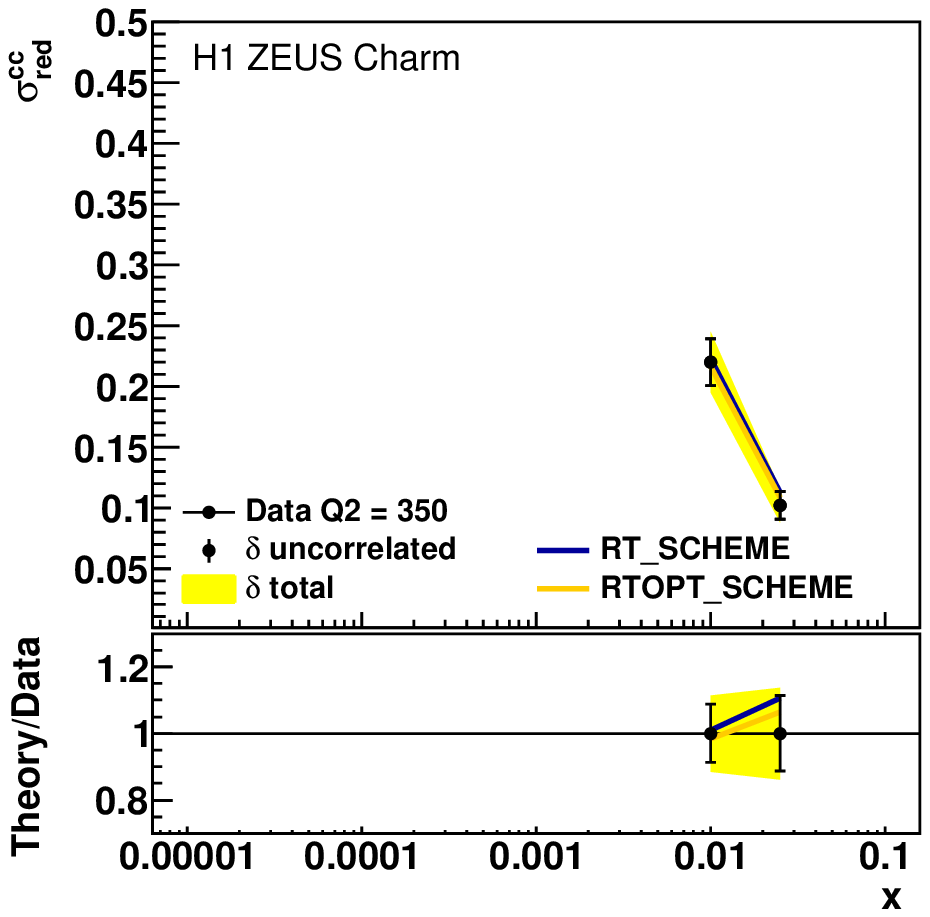}
\includegraphics[width=0.32\textwidth]{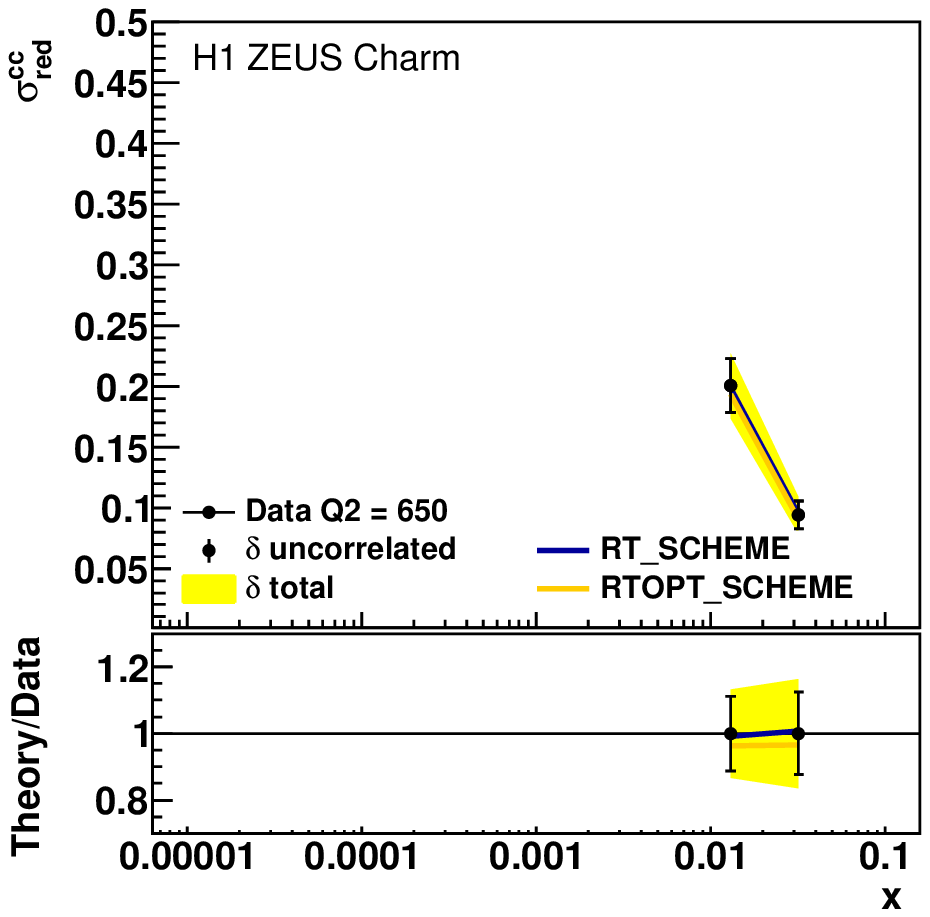}

\includegraphics[width=0.32\textwidth]{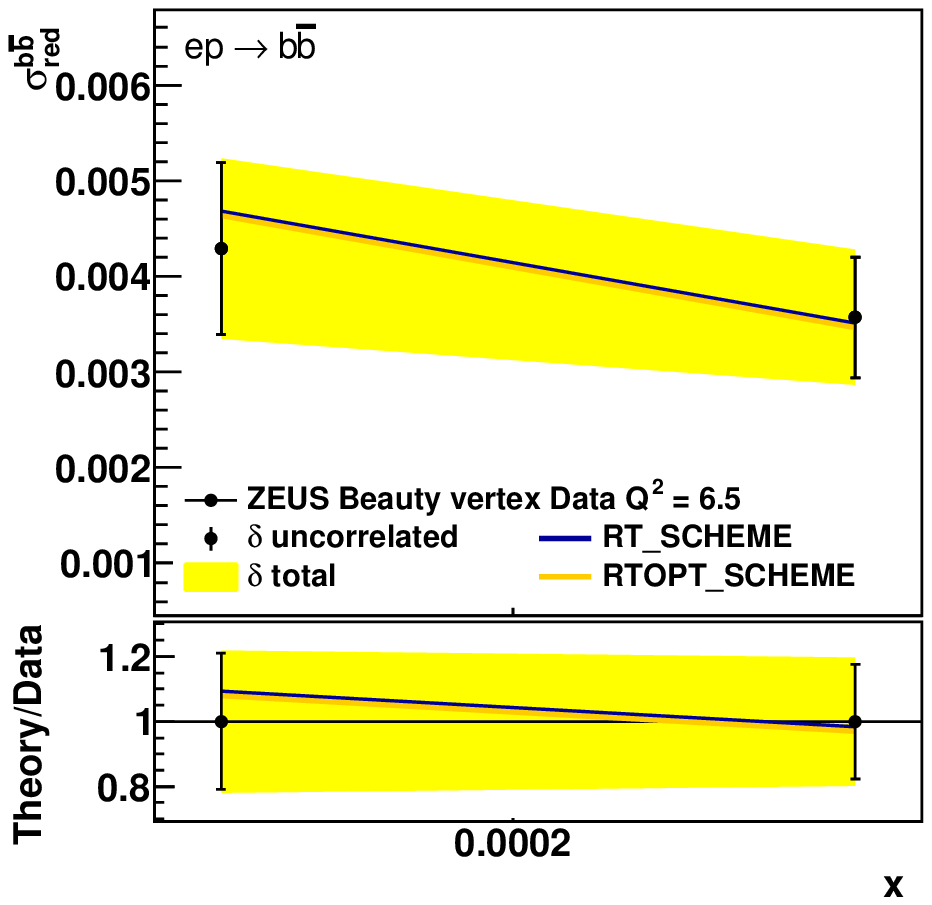}
\includegraphics[width=0.32\textwidth]{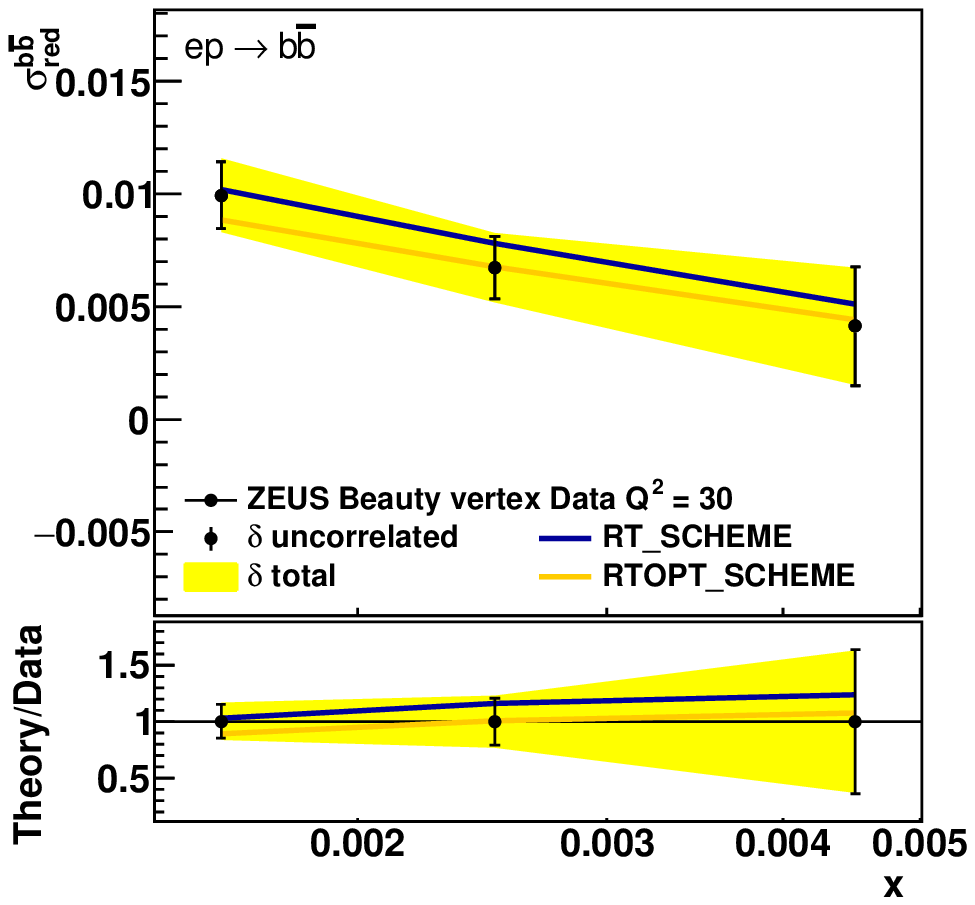}
\includegraphics[width=0.32\textwidth]{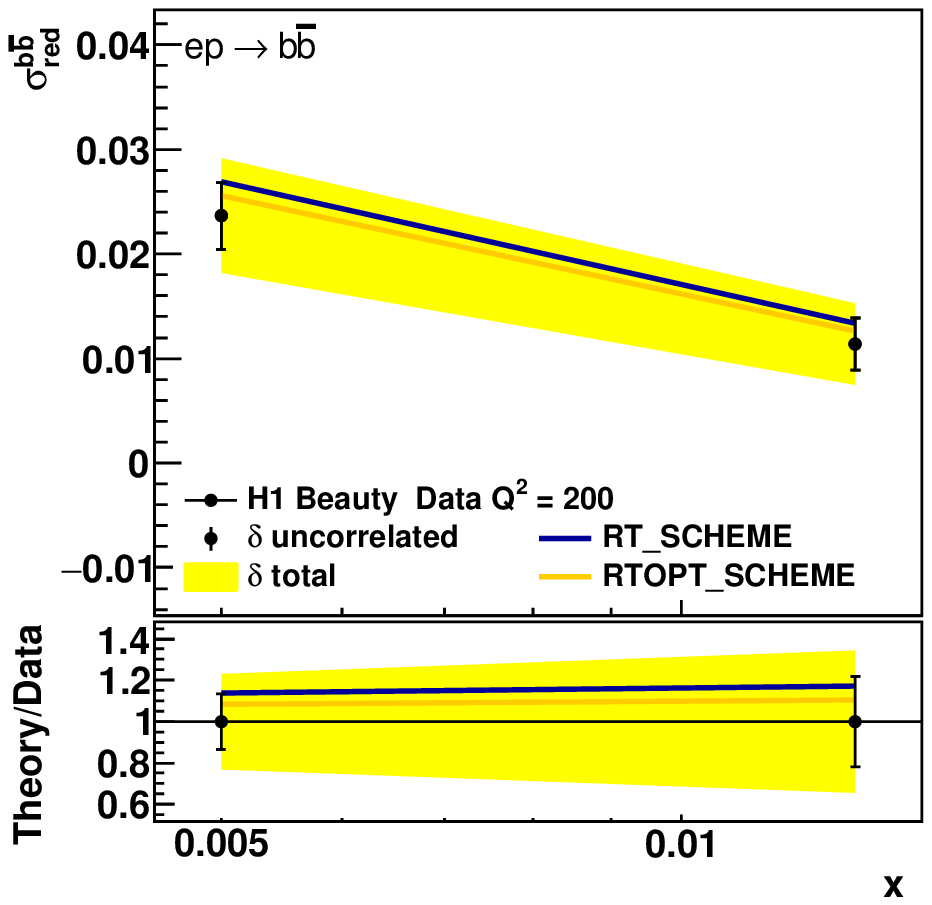}
\caption{Illustrations of the consistency of HERA measurements of the reduced DIS  $e^{\pm}p$ data \cite{Abramowicz:2015mha}, charm and beauty quark cross section  data~\cite{Abramowicz:1900rp,Aaron:2009af,Abramowicz:2014zub}  and the theory predictions as a function of $x$ and for different values of $Q^2$.}
\label{fig:1}
\end{figure*}

 In Fig.~\ref{fig:1},  we illustrate the consistency of HERA measurements of the reduced deep inelastic  $e^{\pm}p$  scattering  cross sections data \cite{Abramowicz:2015mha} and the theory predictions as a function of $x$ and for different values of $Q^2$. According to our QCD analysis, we have a good agreement between the theory and experimental data. The uncertainties on the cross sections in Fig.~\ref{fig:1} are obtained using Hessian error propagation. The corresponding, $\frac{\chi^2_{\rm TOTAL}}{dof}$ values for each of the data sets in Fig.~\ref{fig:1}, are listed in Table~\ref{tab:data}. 
 
\begin{figure*}
\includegraphics[width=0.23\textwidth]{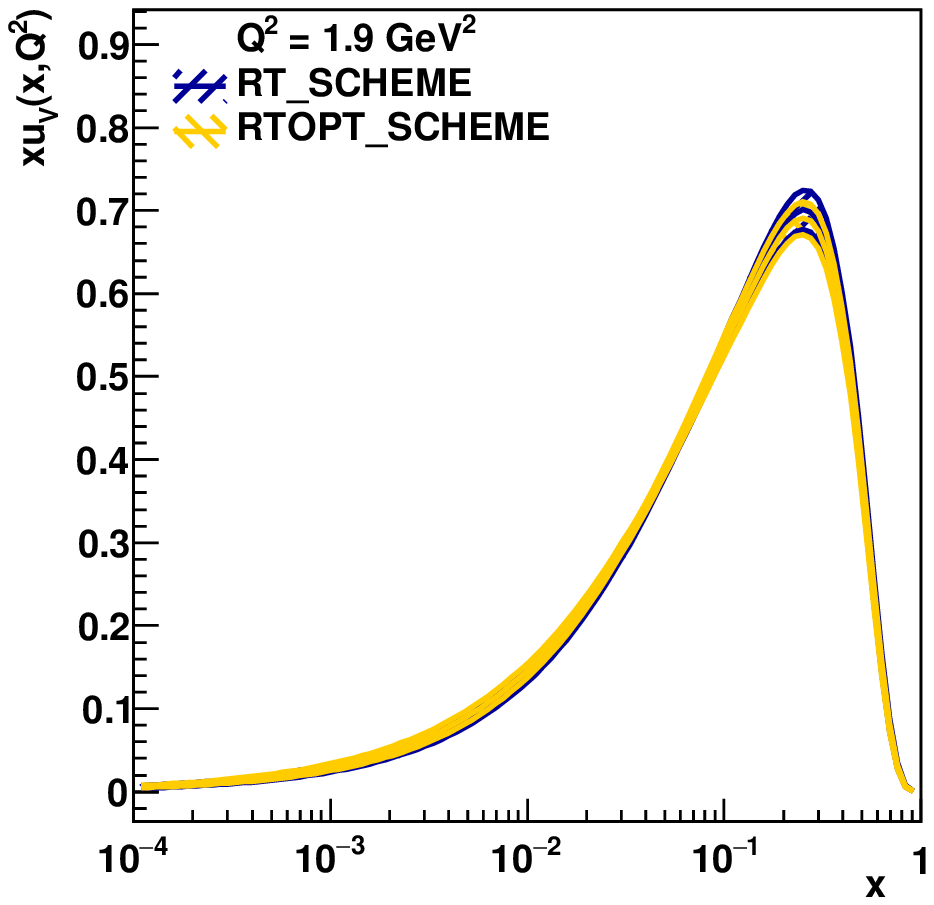}
\includegraphics[width=0.23\textwidth]{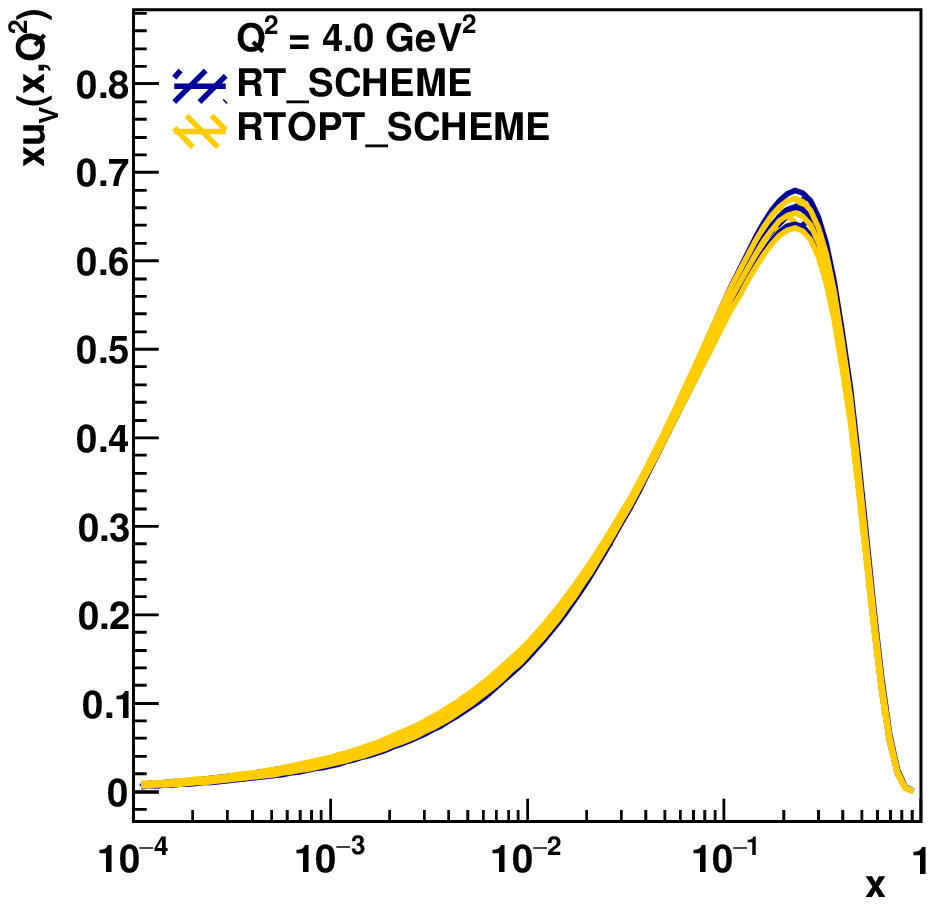}
\includegraphics[width=0.23\textwidth]{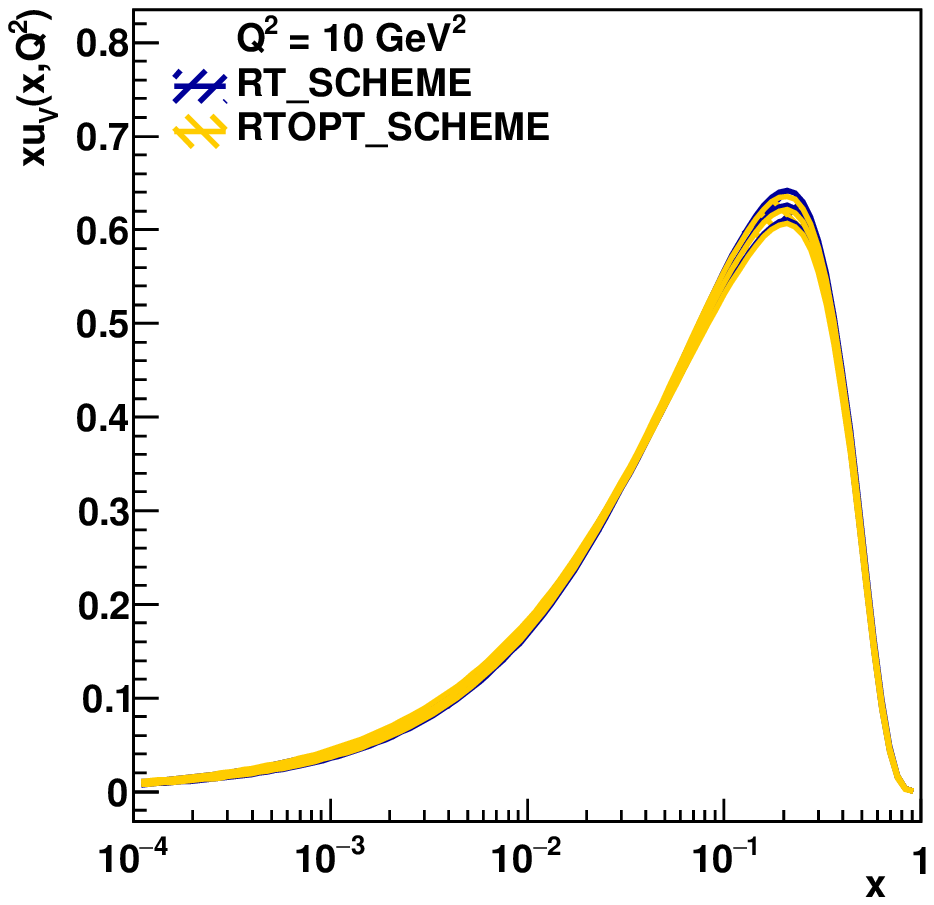}
\includegraphics[width=0.23\textwidth]{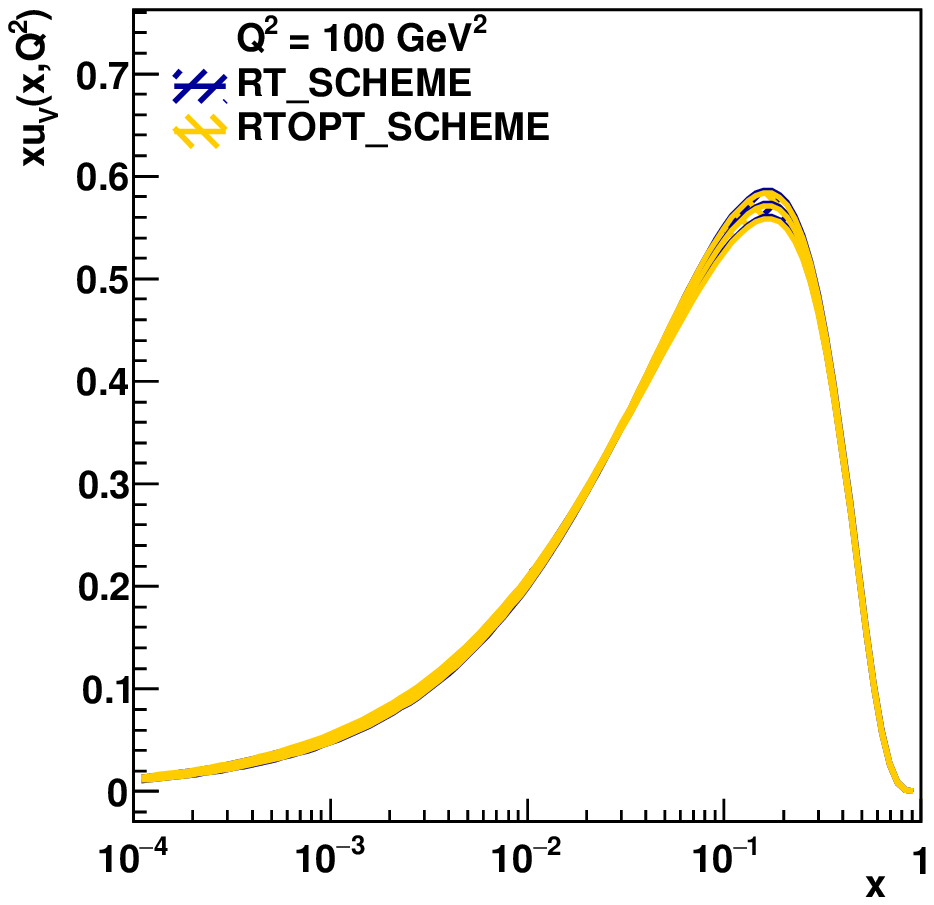}

\includegraphics[width=0.23\textwidth]{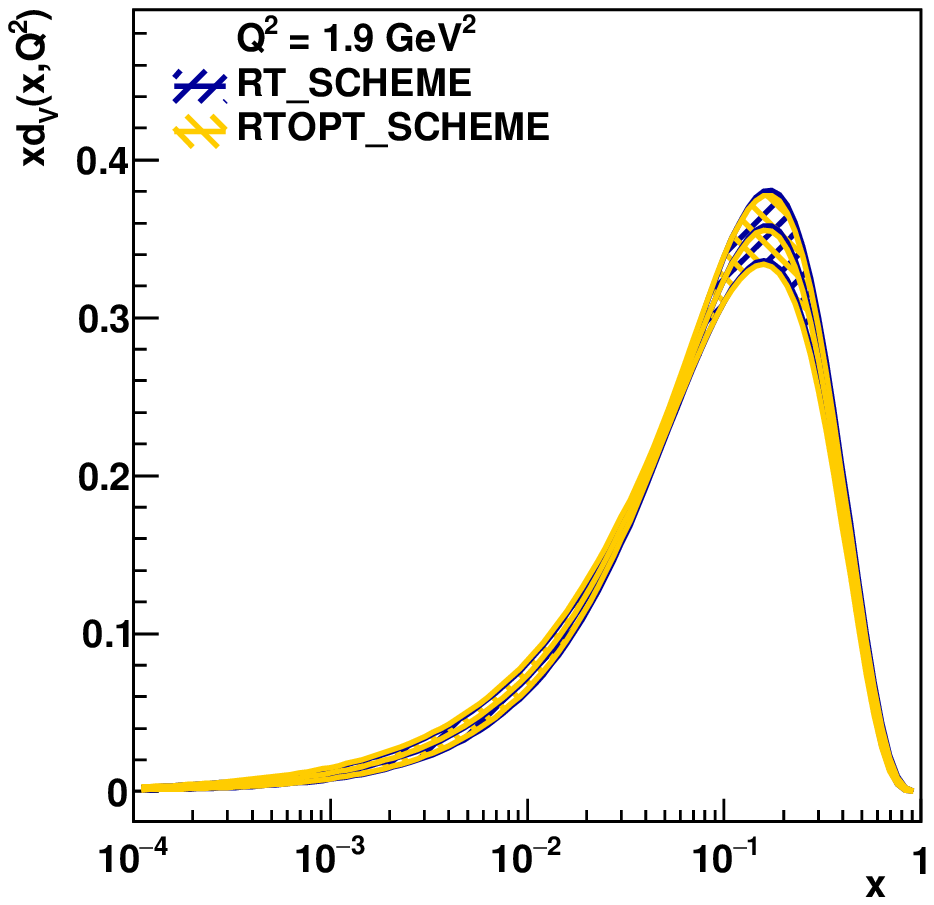}
\includegraphics[width=0.23\textwidth]{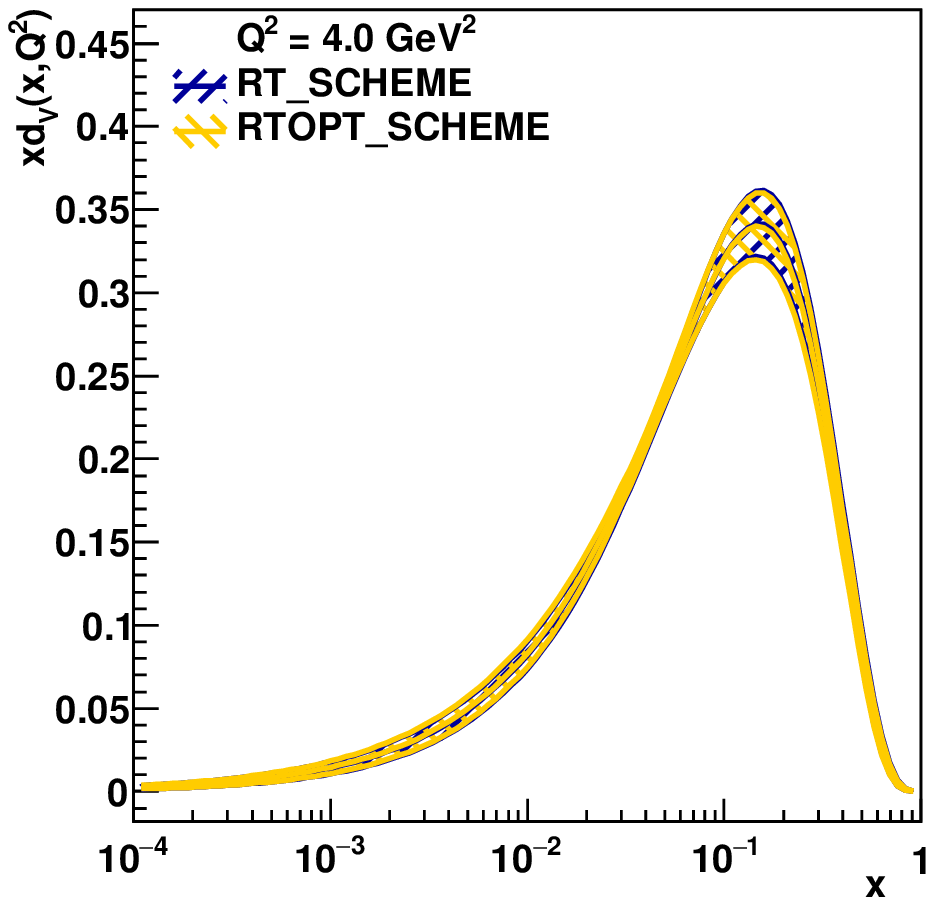}
\includegraphics[width=0.23\textwidth]{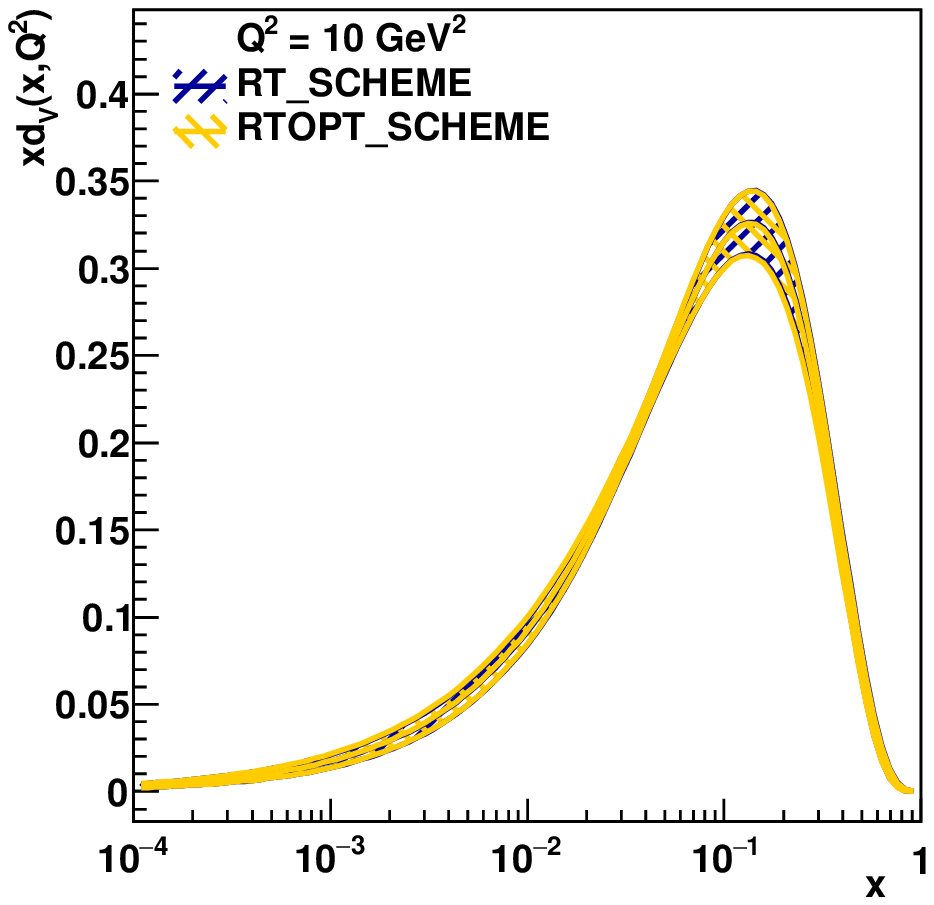}
\includegraphics[width=0.23\textwidth]{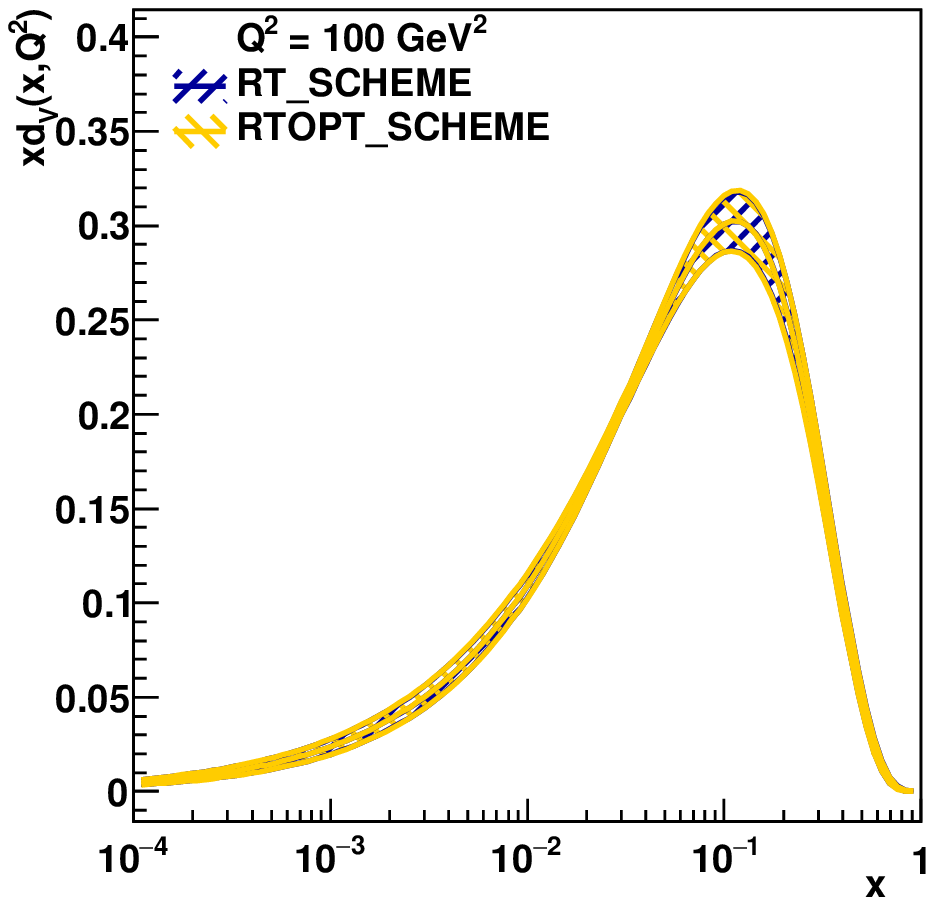}

\caption{The $xu_v$ and $xd_v$ distributions at the starting value $Q_0^2$ = 1.9~GeV$^2$ and $Q^2$ = 4, 10 and 100~GeV$^2$, as a function of $x$. As we can see, the $xu_v$ and $xd_v$ distributions are not sensitive to different schemes.}
\label{fig:2}
\end{figure*}

 In Fig.~\ref{fig:2}, we show $xu_v$ and $xd_v$ distributions at the starting value $Q_0^2$ = 1.9~GeV$^2$ and $Q^2$ = 4, 10 and 100~GeV$^2$, as a function of $x$. On the other hand from our numerical values of $xu_v$ and $xd_v$ PDFs in Table~\ref{tab:parh}, we expect to see no sensitivity of $xu_v$ and $xd_v$ distributions to different schemes. This is consistence with results shown in Fig.~\ref{fig:2}. 

\begin{figure*}
\includegraphics[width=0.32\textwidth]{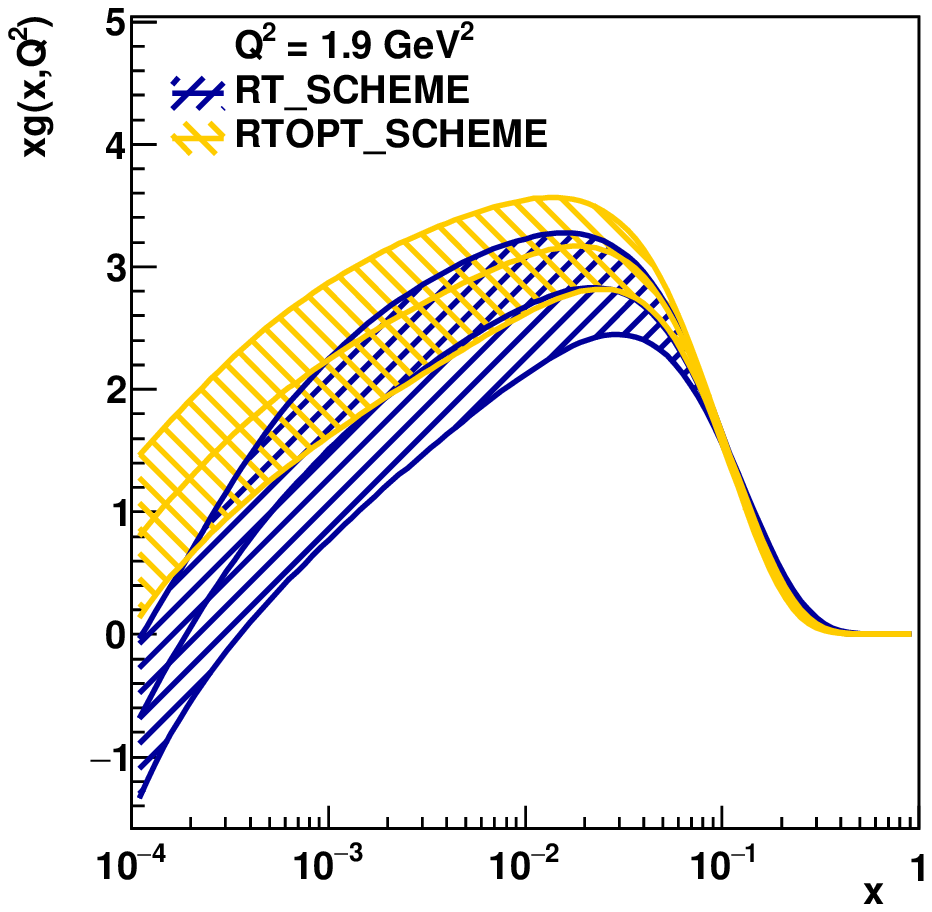}
\includegraphics[width=0.32\textwidth]{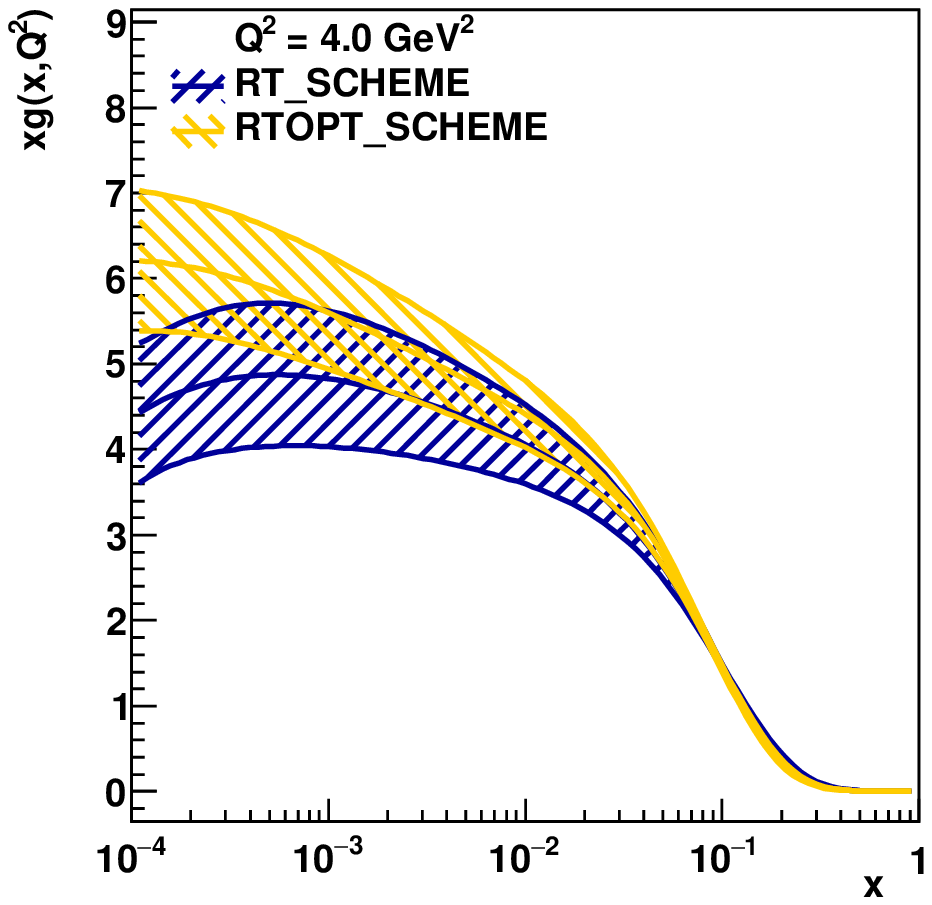}
\includegraphics[width=0.32\textwidth]{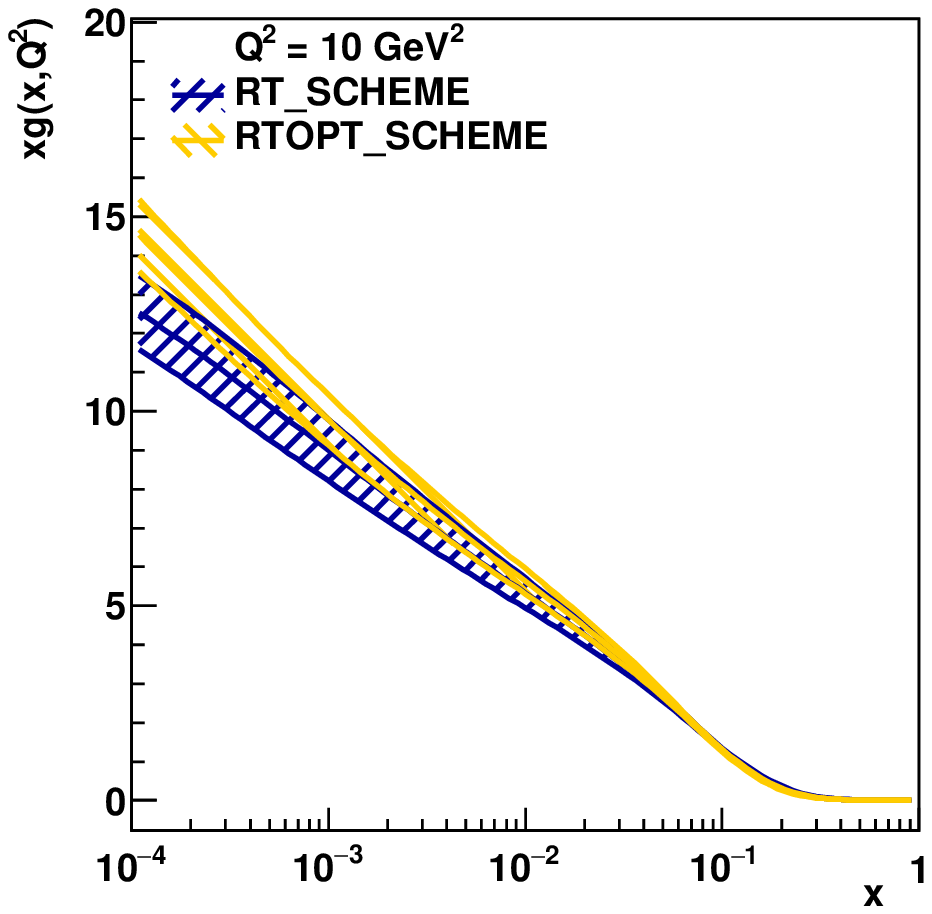}
\caption{The gluon PDFs as extracted for two different RT and RT OPT schemes at the starting value $Q_0^2$ = 1.9~GeV$^2$ and $Q^2$ = 4 and 10~GeV$^2$, as a function of $x$.}
\label{fig:3}
\end{figure*}

\begin{figure*}
\includegraphics[width=0.32\textwidth]{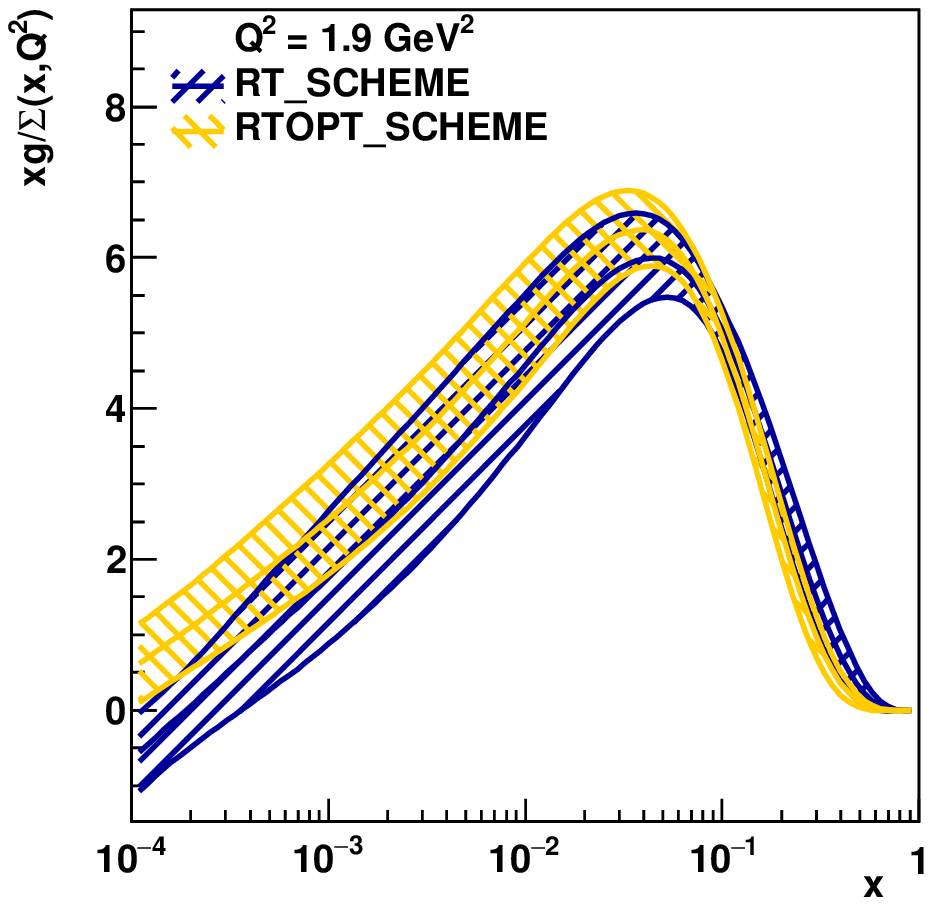}
\includegraphics[width=0.32\textwidth]{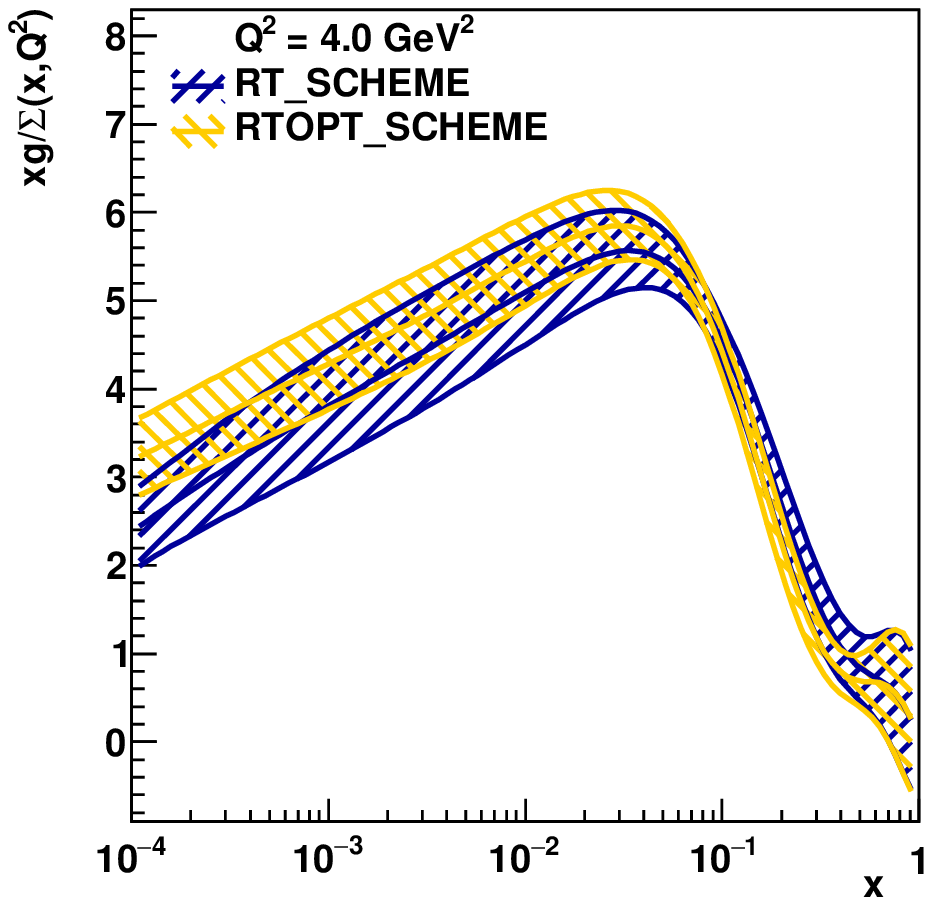}
\includegraphics[width=0.32\textwidth]{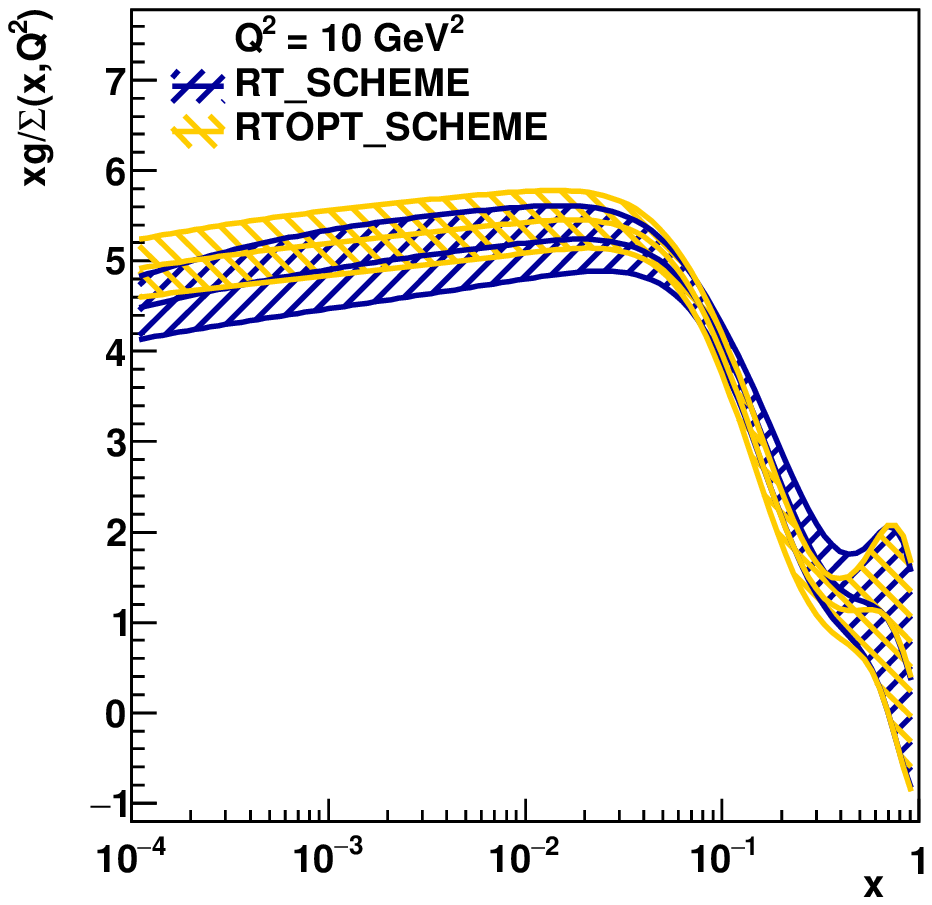}
\caption{The ratio of $xg$ (gluon distribution) over $\Sigma$-PDFs, for two different RT and RT OPT schemes at the starting value $Q_0^2$ = 1.9~GeV$^2$ and $Q^2$ = 4 and 10~GeV$^2$, as a function of $x$.}
\label{fig:4}
\end{figure*}

The gluon PDFs as extracted for two different RT and RT OPT schemes at the starting value $Q_0^2$ = 1.9~GeV$^2$ and $Q^2$ = 4 and 10~GeV$^2$, as a function of $x$ are shown in Fig.~\ref{fig:3}. By having the total sea quark $\Sigma$-PDFs,  defined by $\Sigma=2x(\bar u+\bar d+\bar s+\bar c)$,  
one can plot  the ratio of $xg$ (gluon distribution) over $\Sigma$-PDFs, for two different RT and RT OPT schemes at the starting value $Q_0^2$ = 1.9~GeV$^2$ and $Q^2$ = 4 and 10~GeV$^2$, as a function of $x$. We preset this ratio in Fig.~\ref{fig:4}.

\begin{figure*}
\includegraphics[width=0.32\textwidth]{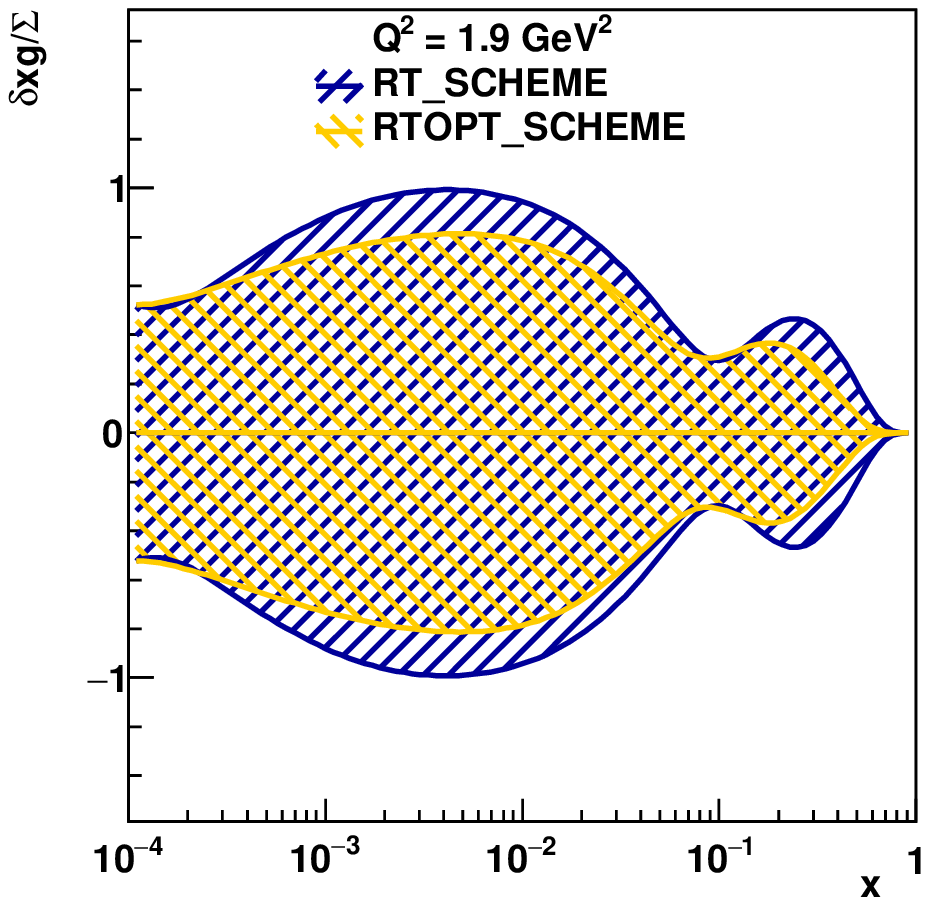}
\includegraphics[width=0.32\textwidth]{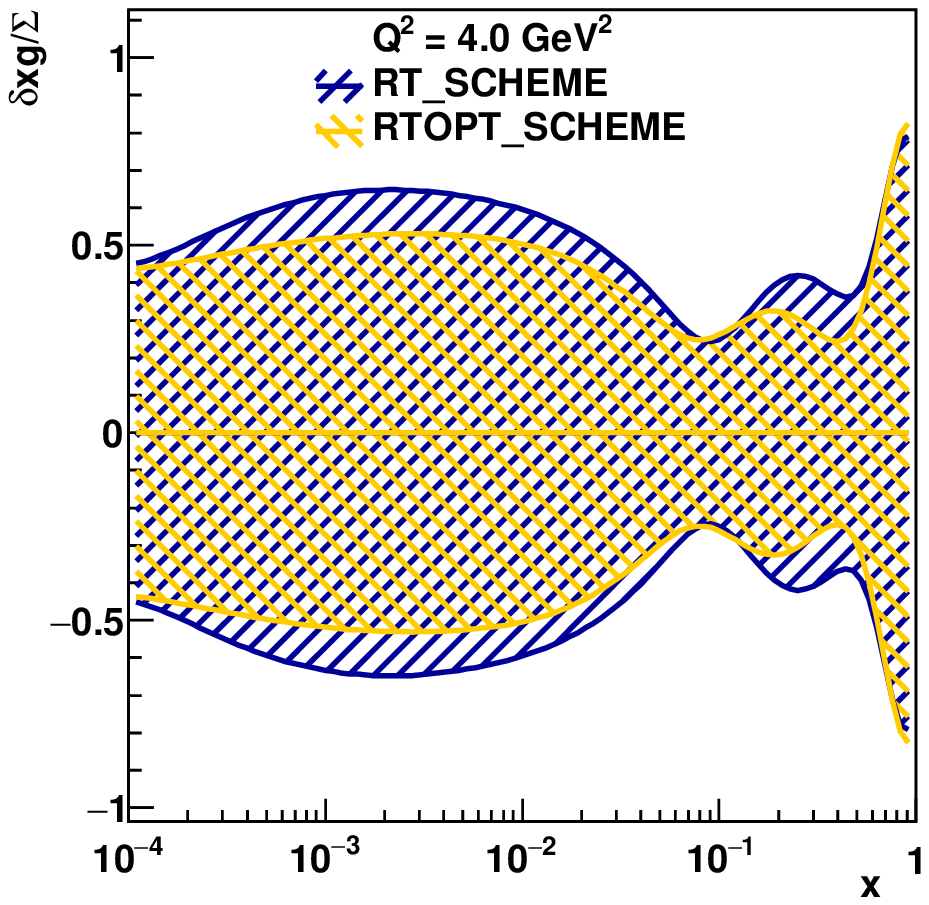}
\includegraphics[width=0.32\textwidth]{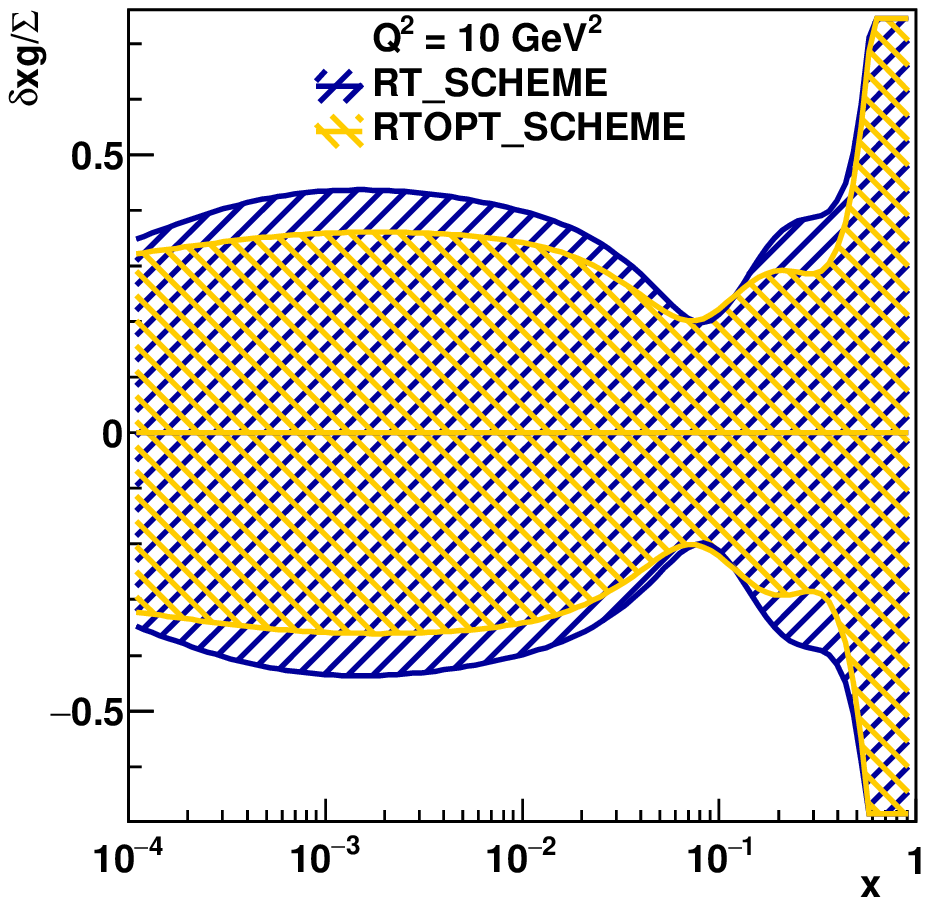}

\includegraphics[width=0.32\textwidth]{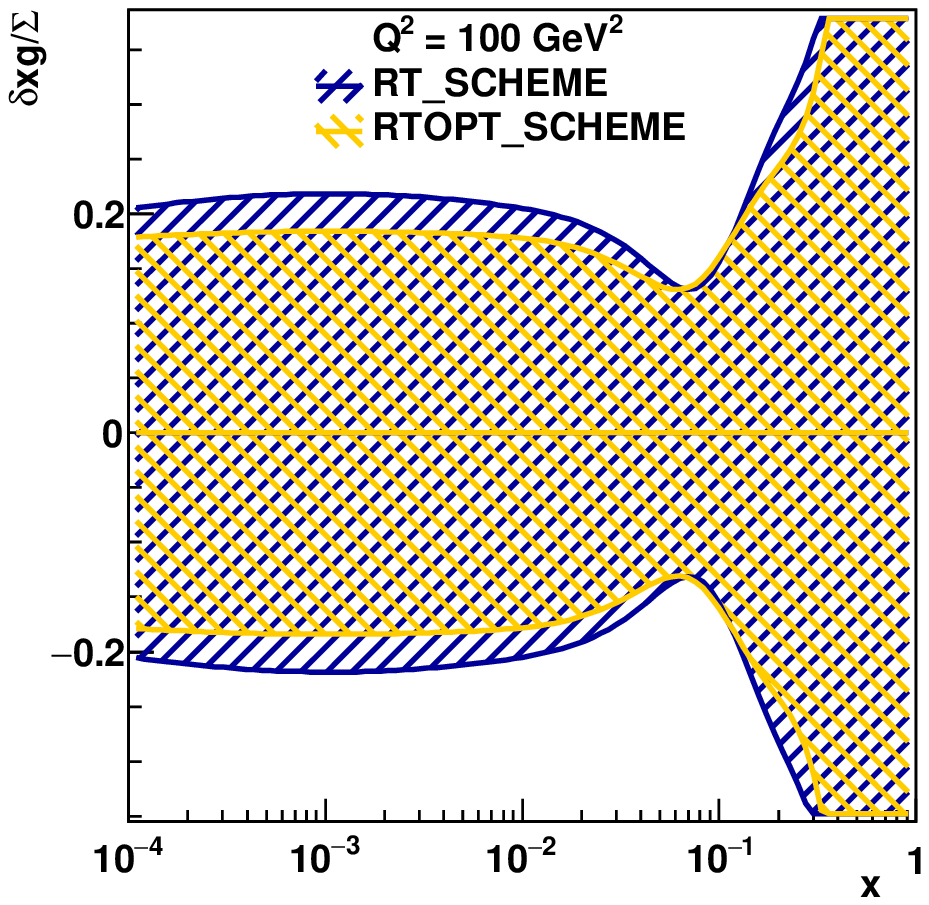}
\includegraphics[width=0.32\textwidth]{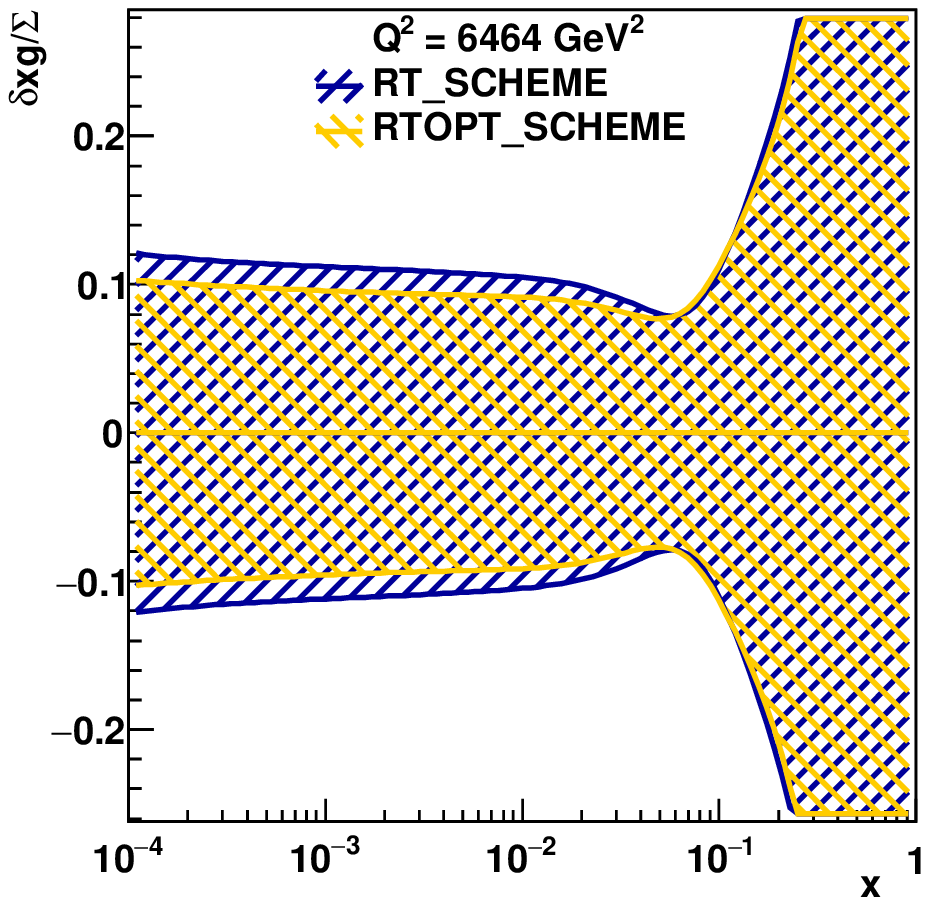}
\includegraphics[width=0.32\textwidth]{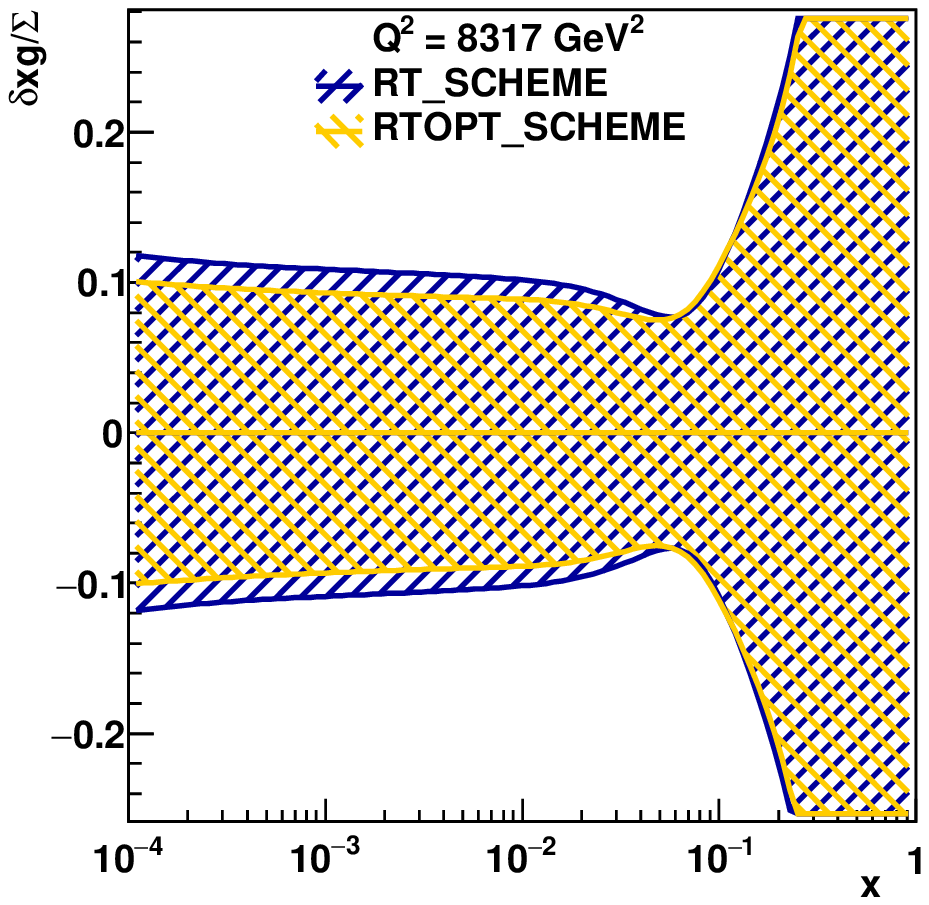}
\caption{The partial ratio of gluon distributions over $\Sigma$-PDFs, for two different RT and RT OPT schemes, at the initial scale $Q_0^2$ = 1.9~GeV$^2$ and $Q^2$ = 4, 10, 100, 6464 and 8317~GeV$^2$, as a function of $x$.}
\label{fig:5}
\end{figure*}

In Fig.~\ref{fig:5}, we present the partial ratio of gluon distributions over $\Sigma$-PDFs for two different RT and RT OPT schemes at the initial scale $Q_0^2$ = 1.9~GeV$^2$ and $Q^2$ = 4, 10, 100, 6464 and 8317~GeV$^2$ as a function of $x$.

The presently  determined ${\alpha_s(M^2_Z)}$ values in our PDF analysis with different data sets and different schemes in the range of $0.1151-0.1177$ is smaller than the PDG world avarge of ${\alpha_s(M^2_Z)}=0.1181\pm0.0013$  which is reported in Ref. \cite{Agashe:2014kda}. Of course, the differences in the values of ${\alpha_s(M^2_Z)}$ from different PDF analysis is due to different data sets used or to different assumptions of theory applied. For example,  the hadro-production  of jets data from the LHC,  have an impact on ${\alpha_s(M^2_Z)}$ value and may provide valuable constraints. In this regards, ${\alpha_s(M^2_Z)}$ measurements are not only depend  on PDFs global fits, but also are depend on different processes and methods at different scales as well.

A standard LHAPDF library file of this QCD analysis at the next-to-leading order is available and can be obtained via e-mail from the authors.

%\section{Acknowledgments}
%We gratefully acknowledge  V. Radescu and R. Placakyte  for guidance and useful discussions about xFitter package. A. K.  is grateful to CERN TH-PH division for the hospitality where a portion of %this work was performed.

\bibliography{avm1b}% Produces the bibliography via BibTeX.

%\clearpage

%\newpage
%\onecolumngrid

\clearpage

\end{document}